\documentclass[reprint,aps,nofootinbib,preprintnumbers,amsmath,amssymb,11pt]{revtex4}
\usepackage{graphicx}
\usepackage{dcolumn}
\usepackage{bm}
\usepackage{natbib}
\usepackage{color}
\newcommand{\nc}{\newcommand}

\newcommand\nn{{\nonumber}}
\newcommand{\beq}{\begin{equation}}
\newcommand{\eq}{\end{equation}}

\newcommand{\half}{\frac{1}{2}}

\nc{\bea}{\begin{eqnarray}} \nc{\ea}{\end{eqnarray}} \nc{\be}{\begin{equation}} \nc{\ee}{\end{equation}} \nc{\barr}{\begin{array}}
\nc{\earr}{\end{array}}

\def\be{\begin{equation}}
\def\ee{\end{equation}}
\def\bea{\begin{eqnarray}}

\def\eea{\end{eqnarray}}

\def\lab{\label}
\def\f{\phi}

\def\e{\epsilon}

\def\le{\left}
\def\ri{\right}

\def\half{\frac12}

\def\cO{{\cal O}}
\def\m{\mu}
\def\n{\nu}

\def\6{\partial}

\def\a{\alpha}

\def\lab{\label}

\def\g{\gamma}
\def\tr{{\rm Tr}}
\def\eps{\epsilon}
\def\la{\langle}
\def\ra{\rangle}
\def\fb{f}

\begin{document}

\title{(Non)renormalization of Anomalous Conductivities and Holography}




\author{Umut G\"ursoy$^1$  and Aron Jansen$^{2}$
\\[0.4cm]
\it $ ^{1,2}$\textit{Institute for Theoretical Physics, Utrecht University\\ Leuvenlaan 4, 3584 CE Utrecht, The Netherlands}}
{\let\thefootnote\relax\footnotetext{$^{1}$U.Gursoy@uu.nl \\ $^{2}$A.P.Jansen@uu.nl}}

\date{\today}

\begin{abstract}

The chiral magnetic and the chiral vortical effects are recently discovered phenomena arising from chiral gauge and gravitational anomalies that lead to generation of electric currents in presence of  magnetic field or vorticity. The magnitude of these effects is determined by the anomalous conductivities. These conductivities can be calculated by the linear response theory, and in the strong coupling limit this calculation can be carried out by the holographic techniques. Earlier calculations  in case of conformal field theories indicate non-renormalization of these conductivities where the holographic calculation agrees with the free field limit. We extend this holographic study to non-conformal theories exhibiting mass-gap and confinement-deconfinement type transitions in a holographic model based on the analytic black hole solution of Gao and Zhang. We show that radiative corrections are also absent in these non-conformal theories confirming indirect arguments of  Jensen et al  in a direct and non-trivial fashion. There are various indications in field theory that such radiative corrections should arise when  contribution of dynamical gluon fields to the chiral anomaly is present. Motivated by this, we seek for such corrections in the holographic picture and argue that such corrections indeed arise through mixing of the background and its fluctuations with the axion and the one-form fields that couple to the flavor and probe gauge branes through the Wess-Zumino terms. These corrections are non-vanishing when the flavor to color ratio $N_f/N_c$ is finite, therefore they are only visible in the Veneziano limit at large $N_c$.       

\end{abstract}

\maketitle
\newpage
\tableofcontents
\newpage

\section{Introduction and Summary}

Anomaly induced transport in systems of chiral fermions is currently a subject of  active theoretical and experimental studies. Various new effects related to the axial anomaly in such quantum field theories were discovered, most notably the Chiral Magnetic Effect, the Chiral Separation Effect and the Chiral Vortical effect.    
The Chiral Magnetic Effect \cite{CME1} provides a macroscopic manifestation of the quantum anomalies. In short, in presence of an external magnetic field $\vec{B}$, a combination of QCD and QED anomalies result in generation of an electric current parallel to $\vec{B}$. There exist various derivations of this phenomenon, directly in perturbative quantum field theory  \cite{CME2} as well as in relativistic anomalous hydrodynamics, \cite{SonSurowka} both resulting in the expression
\be\lab{J1} 
\vec{J} = \frac{e^2}{2\pi^2} \mu \vec{B}\, .
\ee 
 Here the coefficient is the well-known QED anomaly coefficient \cite{ABJ} and $\mu$ is an effective chemical potential representing the imbalance  in the chiral charge, that is generated by non-perturbative processes in QCD which violate chiral charge conservation. At finite temperature in the deconfined phase of QCD, one expects the most dominant such process to be the sphaleron decay \cite{spdecay}.  Anomaly related phenomena can also be studied within relativistic hydrodynamics, when the hydrodynamic approximation  applies. For example, authors of \cite{SonSurowka} (see also \cite{NeimanOz}) presented an independent derivation of (\ref{J1}) by the physical requirement of  non-negative entropy current. In this derivation anomalous transport was generalized to include the effects of relatives of the Chiral Magnetic Effect (CME)  such as the Chiral Vortical Effect (CVE), that is, generation of an electric current in the presence of vorticity due to the gravitational anomaly \cite{Erdmenger, Banerjee}. Similarly, the chiral and the gravitational anomalies also give rise to anomalous heat transport in the presence of magnetic field and vorticity respectively. On the experimental side, all of these phenomena  can, in principle, be realized  in the Heavy Ion Collision experiments, although experimental evidence  is controversial at present \cite{Exp}.   

Agreement of the perturbative and the hydrodynamic calculations imply non-renormalization of the chiral magnetic conductivity coefficient in (\ref{J1}). However, as we explain below this issue is more subtle\footnote{We thank Karl Landsteiner and Amos Yarom for very useful discussions on the current situation of (non)renormalization of anomalous conductivities.}. In this paper we address the question of renormalization of chiral magnetic and vortical conductivities in electric and heat currents and related in the holographic approach in a gravitational setting dual to a non-conformal, confining QFT at finite temperature\footnote{It is worth-mentioning the historical fact that in case of the chiral vortical effect, such anomalous transport was first discovered in the context of holography\cite{Erdmenger, Banerjee} which then triggered a direct hydrodynamics investigation.}.


Anomalous transport coefficients associated with the electric and heat currents are non-dissipative and can also be calculated by use of linear response theory (see for example the review \cite{LandsteinerR}.) at vanishing frequency. With no loss of generality one can consider turning on a gauge and metric fluctuations $\delta A_z(k_y)$ and $g_{0z}(k_y)$ to introduce magnetic field $B_x$ and vorticity $\omega_x$ and measure the response  $\la J^x\ra$ and $\la T^{0x} \ra$ obtaining the associated transport coefficients as 
         
\begin{equation}\lab{kubo}\begin{split}
\sigma_B &= \lim_{k_y \rightarrow 0} \frac{i}{k_y} \langle J^x J^z \rangle \, , \\
\sigma_V &= \lim_{k_y \rightarrow 0} \frac{i}{k_y} \langle J^x T^{0 z}\rangle \, , \\
\sigma_B^\epsilon &= \lim_{k_y \rightarrow 0} \frac{i}{k_y} \langle T^{0x} J^z\rangle  \, , \\
\sigma_V^\epsilon &= \lim_{k_y \rightarrow 0} \frac{i}{k_y} \langle T^{0x} T^{0z}  \rangle \, .
\end{split}\end{equation}
Here $\sigma_B$,  $\sigma_V$, $\tilde{\sigma}_B$ and $\tilde{\sigma}_V$ denote the chiral magnetic and chiral vortical conductivities in the electric and the heat currents respectively. $\sigma_V$ and $\tilde{\sigma}_V$ are equal due to symmetry of the two-point functions at zero frequency. At strong coupling these quantities should be calculated using non-perturbative techniques such as the holographic correspondence. 

This paper is concerned with quantum corrections of these transport coefficients at strong coupling. The main question we address is whether the conductivities (\ref{kubo}) renormalize or not in presence of strong interactions. There exist a variety of arguments (proofs in certain cases) in favor of---at least perturbative---non-renormalization \cite{NonRenorm, CME2, JensenNR, Minwalla, Roy, Safodyev1, Safodyev2}, mainly due to the fact that the anomaly coefficients are one-loop exact \cite{ABJ}, and one expects to be able to prove this directly in QFT by using the anomaly equations and relevant Ward identities. However, there also exist calculations and arguments in favor of renormalization in certain cases \cite{Renorm}. 
It is fair to say that  the matter turns out to be sufficiently complicated to provide a direct proof within traditional QFT, and indeed  such a proof including also non-perturbative corrections does not exist\footnote{The most clear situation is the case of the CME coefficient in field theories where dynamical glue fields do not contribute to the anomaly. In these theories Ward identities and the anomaly equations are sufficient to fix the CME coefficient, see for example \cite{JensenNR} for a direct proof for theories with finite static correlation length. }. We refer the reader to section 5 of \cite{Buividovich} for a clear account of the current situation. In the case of the chiral magnetic conductivity there exist various direct field theory calculations using the axial and vector Ward identities and 
some recently proven non-renormalization theorems \cite{NonRenTheorem}. These arguments however may not be applicable  at finite temperature, and they also ignore non-perturbative contributions. In the case of Chiral Vortical Conductivity, there exist no such direct field theory argument against quantum corrections at finite temperature\footnote{See \cite{Roy} for an alternative proof of non-renormalization based on a group theoretic analysis at zero temperature}.  
On the contrary, both field theory calculations \cite{GolkarSon, HouRen}  and lattice simulations \cite{lattice} indicate renormalization effects in the chiral vortical  conductivity. 

The question was considered in the holographic dual description \cite{Malda, Polyakov, Witten1} in the special case of ${\cal N}=4$ super Yang-Mills conformal plasma in the large $N_c$ limit in a series of papers by Landsteiner et al. \cite{Landsteiner1, Landsteiner2, Landsteiner3, Landsteiner4}. By comparison of the holographic and weak coupling results, these authors  concluded that none of the transport coefficients receive quantum corrections hence obtaining a puzzling result in view of the previous paragraph. The  conductivities in (\ref{kubo}) are found to be: 
\begin{equation}\label{J2}\begin{split}
\sigma_B &=  \frac{\mu }{4 \pi^2} \, , \\
\sigma_V &=  \left( \frac{\mu^2}{8\pi^2} +  \frac{T^2}{24} \right) \, , \\
\sigma_B^\epsilon &=   \left(\frac{\mu^2}{8\pi^2} + \frac{T^2}{24} \right) \, , \\
\sigma_V^\epsilon &= \frac{\mu^3}{12 \pi^2}+ \frac{1}{12} \mu T^2 \, .
\end{split}\end{equation}
There are a variety of reasons to believe that the aforementioned holographic calculation is too specific to answer the question in full generality. Firstly it requires infinite 't Hooft coupling $\lambda$ and infinite $N_c$, rendering possible corrections in $1/\lambda$ and $1/N_c$ invisible.  Secondly, it applies to  conformal plasmas where the only dimensionful parameters are $T$ and $\mu$. In a theory such as QCD, there exists a dynamically generated scale $\Lambda_{QCD}$ that arises from dimensional transmutation. Finally, QCD is a confining theory with  mass gap and confinement-deconfinement transition (cross-over) at a temperature $T_c$ both proportional to $\Lambda_{QCD}$. One can easily imagine that dynamics that lead to confinement at low T, placing the theory in a different universality class than conformal theories, may lead to a different result\footnote{This question was addressed in the holographic setting of the soft-wall model\cite{SoftWall} in \cite{Gorsky}. However we believe the soft-wall model is not  appropriate to address the question because the holographic calculation makes use of fluctuating background fields, whereas the soft-wall model does not even provide a genuine solution to the Einstein's equations.}. Indeed a lattice calculation of the chiral vortical conductivity shows non-trivial dependence on temperature in the confining versus deconfined phases of QCD-like theories\cite{lattice}.  

Finally, in a beautiful paper by Jensen, Loganayagam and Yarom \cite{Yarom} (see also \cite{Yarom2}) it was argued that when the hydrodynamic description is valid, one expects no renormalization neither in the chiral magnetic nor in the chiral vortical conductivities regardless of whether the underlying theory is conformal or not. The arguments in \cite{Yarom} are based on placing the theory on a cone, constructing  the Euclidean generating function and requiring continuity of this generating function in the limit where the deficit angle vanishes. There are various reasons however to believe that quantum corrections would arise when some of the assumptions in this derivation are lifted. One such case is considered in \cite{Kovtun} (see also \cite{Banerjee2}) where the authors make the distinction between type I and type II anomalies. The former vanishes when the external fields are turned off, whereas the latter does not. The gluonic contribution is of type II in this classification and the arguments in \cite{Yarom} does not apply to this case\footnote{We thank Amos Yarom for pointing out these references to us.}.  
We further discuss these issues  in section \ref{gluonsection}. The bottom-line of all of this discussion is that  it is of considerable interest to provide a direct check of the arguments in \cite{Yarom} and address the question of non-renormalization in an independent manner.   

In this paper we address the question in the holographic setting dual to a non-conformal plasma with a mass gap, sharing many of the salient features of a gravity background dual to QCD at finite T. We consider a bottom-up approach to holography and perform our calculations in a black-brane background that is an asymptotically AdS solution to the Einstein-Maxwell-dilaton system in 5 dimensions.  Dynamics of the dilaton is determined by a non-trivial dilaton potential that makes the dilaton run as a function of the holographic coordinate generating a mass scale $\Lambda_{QCD}$ in the dual field theory.  For technical reasons it is very helpful to have analytic solutions at hand  and  for this reason we consider a specific dilaton potential constructed first in \cite{GaoZhang} by Gao and Zhang that leads to such analytic backgrounds. 

We introduce the background of \cite{GaoZhang} and derive its thermodynamic properties in section \ref{Background}. The dilaton potential depends on an adjustable real parameter $\alpha$.  The particular choice of $\alpha=0$ precisely corresponds to the aforementioned ${\cal N} =4$ case, whereas for non-vanishing values of $\alpha$ generically corresponds to a non-conformal dual theory. For the particular value of $\alpha=2$ we show that the solution admits a Hawking-Page type transition that is believed to correspond to the confinement-deconfinement transition in the dual field theory. Therefore the backgrounds we consider in this paper encompass all  cases of conformal, non-conformal and confining theories.  Details of this background is presented in the section \ref{Background}. We introduce the anomalies of type I in the language of \cite{Kovtun} in this holographic setting through the AFF and ARR Chern-Simons terms in 5D\footnote{The ARR Chern-Simons term corresponds to a mixed gauge-gravitational anomaly in the dual field theory. We introduced this term following the conjecture of \cite{Landsteiner3} which relates the origin of the $T^2$ term in the axial vortical effect to mixed gauge-gravitational anomaly. It is shown in \cite{Kalaydzhyan} that this conjecture does not represent the full generality of the situation. We thank Tigran Kalaydzhyan for driving our attention to this point.}.       

We calculate the conductivities  (\ref{J1}) in this theory in section \ref{main} as a function of the ``non-conformality parameter'' $\alpha$ in section 3. The background, the fluctuation equations, the temperature T and the chemical potential $\mu$ all depend non-trivially on $\alpha$. Yet, when the $\sigma$'s are expressed in terms of $\mu$ and $T$ in this background, we  find that the form given in (\ref{J2}) holds independently of the value of $\alpha$. From a technical point of view this happens in a very non-trivial manner as a result of delicate cancellations. In this section we also provide a confirmation of these results in an independent manner using the so-called ``holographic flow equations'', generalizing the calculation of \cite{HolFlow} to the non-conformal case.   

While the calculations presented in section \ref{main} provide a non-trivial check of the general result of \cite{Yarom}, they also suggest a way to go around them. In the final section of this paper we present a modification of the holographic calculation that violates the assumptions in \cite{Yarom} resulting in possible quantum corrections to the all of the anomalous conductivities (\ref{J1}). Indeed, the calculation of \cite{GolkarSon} shows that the only possible quantum corrections can arise from contribution of dynamical gauge fields to the loop diagrams. Both in the holographic approach and in the hydrodynamic construction of \cite{Yarom} one works in a limit where the glue fields are non-dynamical. One manifestation of this simplification is the fact that the $U(1)$ axial anomaly in these approaches are given by\footnote{We shall consider only left-handed chiral fermions transforming in the fundamental representation of flavor group $SU(N_f)$ and gauge group $SU(N_c)$ in this paper, although our results are trivially generalizable to other cases.}  
\begin{equation}\label{axanom1}
D\cdot J^{5} = N_c \, a_1\, \tr (F \cdot \tilde{F} )+ N_c \, a_2\, R \cdot \tilde{R} \, , 
\end{equation}
where $F$ is the electromagnetic field strength, $R$ is the Riemann tensor, $\tilde{F}^{\mu\nu} = \epsilon^{\mu\nu\rho\sigma} F_{\rho\sigma}$ and $R \cdot \tilde{R} = \epsilon^{\mu\nu\rho\sigma} R^\alpha{}_{\beta\mu\nu} R^\beta{}_{\alpha\rho\sigma}$,
instead of the full QFT result\footnote{Here $a_1$, $a_2$ and $a_3$ are just numerical factors independent of $N_c$ and $N_f$, whose value depend on the charge of the fundamental representation and whether $J^5$ denote a ``consistent'' or ``covariant'' current.}  
\begin{equation}\label{axanom2}
D\cdot J^{5} = N_c \, a_1\, \tr(F \cdot \tilde{F}) +  N_c \, a_2\,  R \cdot \tilde{R} + N_f\, a_3\, \tr (G \cdot \tilde{G} )\, , 
\end{equation}
with $G$ the gluon field strength and $\tilde{G}$ defined similarly as $\tilde{F}$. 
Therefore, in order to include such corrections in the holographic approach, one should first consider correcting the axial anomaly equation (\ref{axanom1}). The question of generating gluon field contribution to the chiral $U(1)_R$ anomaly in the holographic setting was first addressed by Ouyang, Klebanov and Witten in the context of  ${\cal N}=1$ cascading $SU(N+M)\times SU(N)$ gauge theory in \cite{OKW}. The authors showed that the anomaly arises as a result of non-invariance of the $C_2$ form in the dual background. In effect, presence of a non-trivial $C_2$ generates a mass term for the gauge field dual to $U(1)_R$ current and provides the anomalous contribution\footnote{The same problem was studied in the context of M-theory in \cite{GHP}.}.  The question was addressed in more generality in \cite{Paredes} where the gluonic correction in (\ref{axanom2}) is argued to arise from a  p-form field in the background that couples to both the flavor branes and the probe branes. In the case of interest in this paper, namely in the 5D setting of the bottom-up approach, the role played by the $C_2$ in \cite{OKW} is played by the axion field $C_0$. 

In section \ref{gluonsection} we extend the calculation to include $N_f$ number of flavors through space-filling D4 branes and include in the calculation the $C_0$ field, among other relevant form-fields. We show that\footnote{See also \cite{Iatrakis}.} the axion $C_0$ comes with an addition to the action of the form 
\be\lab{C0act} 
S_a \sim \int d^5x \sqrt{g} Z_0(\f) \le(dC_0 - N_f w_2\, A \ri)^2\, ,
\ee 
where $A$ is the dual of the chiral current and $Z_0$ is  some functions of the dilaton, $N_f$ is the number of flavors and $w_2$ is a constant. This addition on one hand corrects the anomaly equation (\ref{axanom1}) into  (\ref{axanom2}), on the other hand it changes the fluctuation equations of the gauge fields $\delta A_x$ and $\delta A_z$ in the calculation of (\ref{kubo}) because of the mass term.  
Therefore {\em we propose inclusion of the axion field as the holographic mechanism to  generate renormalization of the anomalous conductivities that is expected to arise in presence of dynamical gauge fields.} 
The effective mass term for the gauge field in (\ref{C0act}) turns out to be order $N_f/N_c$, therefore we expect corrections anomalous conductivities to be of order $N_f/N_c$. We conclude that they should be visible only in the Veneziano limit where $N_f/N_c$ is kept finite. We do not attempt at calculation of such corrections in this paper, postponing the study in a future work.        


We end the paper by discussing the results obtained in various different approaches to anomalous transport and  possible applications and extensions of our work in section \ref{Discussion}. Appendices A to E detail our calculations.


\section{The Gravitational Background}
\lab{Background}

We work in the bottom-up approach to holography in this paper and consider black-hole solutions to an Einstein-Maxwell-dilaton theory in 5D with the action
\be\lab{action}
S=-\frac{1}{16\pi G }\int d^{5}x\sqrt{-g}\left( {R}-\frac43(\nabla \Phi
)^{2}-V(\Phi )%
-Z(\f) F_{\mu \nu }F^{\mu \nu }\right)  + \frac{1}{8\pi G }\int_{\6 M} d^4x \sqrt{h} K\, ,  
\ee
where $V(\Phi )$ is a potential term for the
dilaton field $\Phi $, and there is a non-minimal coupling between the 
dilaton and and the electromagnetic field specified by the function $Z(\f)$.  The second term is the Gibbons-Hawking term on the boundary.\footnote{We do not need to add a  counterterm action in this section where we consider  the thermodynamic properties of the system by evaluating the difference between the on-shell black-hole and thermal gas actions. The counter terms  cancel in the difference.}. 

\subsection{An analytic black hole solution} 

In \cite{GaoZhang} the authors found the  following analytic solution to (\ref{action}) for the following specific choices of the potentials:  
\be
{V}({\Phi }) =-\frac{3}{(2+\alpha ^{2})^{2}}\Bigg\{4\alpha ^{2}(\alpha^2-1)%
 e^{-\frac{8\Phi}{3\alpha }}
+4(4-\alpha ^{2})e^{\frac{4\alpha \Phi}{3}}+24\alpha
^{2}e^{-\frac{2(2-\alpha^2)\Phi}{3\alpha }}\Bigg \}\, ,
\label{V}
\ee
and
\be\lab{Z} 
Z(\f) = e^{-\frac43\alpha \Phi}\, .
\ee
We note that the $\alpha =0$ corresponds to the
usual Einstein-Maxwell-dilaton theory, with a constant potential $V= -12$. 

 Expanding the dilaton potential (\ref{V}) near $\f=0$ one finds that 
\be\lab{Vbnd} 
V(\f) = V_0 + \half m^2 \phi^2 + \cdots 
\ee 
where 
\be\lab{Vbnd2}
V_0 = -12, \qquad m^2 = -\frac{32}{3}\, .
\ee
{\em We emphasize that $m^2$ is independent of $\a$.} This  mass term precisely saturates the Breitenlohner-Freedman bound\cite{BF}  and corresponds to a deformation in the boundary theory by VeV of an operator of scale dimension 2. We thus learn that conformal symmetry in the dual field theory is spontaneously broken in the UV. Therefore for any value of $\alpha\neq 0$ the dual field theory is non-conformal. An analytic black hole solution for arbitrary $\a$ can be found \cite{GaoZhang} (see also \cite{Iranian}) as\footnote{We follow the notation of \cite{Iranian}},

\begin{equation}\lab{met2}
d{s}^{2}=-N^{2}(r)\fb^{2}(r)dt^{2}+\frac{r^{2}dr^{2}}{(r^{2}+b^{2})\fb^{2}(r)}%
+(r^{2}+b^{2}){R^{2}(r)}d\Omega _{n-1}^{2}\, ,
\end{equation}
where the coordinates $r$ assumes the values $0\leq r<\infty $, and $%
N^{2}(r) $, $\fb^{2}(r)$, $\Phi (r)$ and $R^{2}(r)$ are given as
\begin{eqnarray}  \label{gr}
N^{2}(r) &=&\Gamma ^{-\gamma},  \label{Nr} \\
\fb^{2}(r) &=&\frac{r^{2}+b^{2}}{l^{2}}\Gamma ^{2\gamma }-
\frac{c^2}{r^{2}+b^2} \Gamma ^{1-\gamma },
\label{fr} \\
\f(r) &=&\frac{3}{4}\sqrt{\gamma (2-2\gamma)}\log \Gamma ,
\label{Phir} \\
R^{2}(r) &=&\Gamma ^{\gamma },  \label{Rr} \\
\lab{Gamma}
\Gamma &=&\frac{r^2}{r^{2}+b^{2}}\, .  
\end{eqnarray}
with
\be\lab{gamma}
\g = \frac{\a^2}{2+\a^2}, \qquad 0\leq \g \leq 1\, .
\ee
Location of the horizon $r_h$ is determined  by $f(r_h)=0$ and related to the integration constants $b$ and $c$ above as 
\be\lab{ceq}
c = r_h^{3\g-1} (r_h^2 + b^2)^{\frac32(1-\g)}\, .
\ee
The field strength and the corresponding electromagnetic potential reads 
\begin{equation} 
F_{rt}=\frac{Qr}{\left( r^{2}+b^{2}\right)^2}, \qquad A_{t}=\mu-\frac{Q}{2(r^{2}+b^{2})}\, .
\label{Ft}
\end{equation}
Regularity at the horizon $r_h$ then determines 
\be\lab{qrel2} 
Q = 2\mu (r_h^2 + b^2) \, .
\ee
The VeV of the dilaton operator dual to $\f$ can be read off from the near boundary asymptotics of (\ref{fr}) as 
\be\lab{VeV}
\la {\cal O} \ra = \frac{3}{4}\sqrt{\gamma (2-2\gamma)}\, b^2\, .
\ee
Therefore, one can think of the integration constant $b$ as related to the dynamically generated mass scale in the dual theory, i.e.  $\Lambda_{QCD} \propto b$. The temperature is obtained by requiring absence of a conical singularity at the horizon as, 
\be
\lab{T2}
T =  \frac{b}{\pi}\,r_h^{3\g-1} (r_h^2 + b^2)^{\half(1-3\g)}\le( \frac{r_h}{b} + \frac{3\g-1}{2}\frac{b}{r_h}\ri)\, .
\ee
Entropy density of black-hole is determined from the area of the horizon as, 
\be
\lab{S2}
S = \frac{r_h^{3\g} (r_h^2+b^2)^{\frac32(1-\g)}}{4G} \, .
\ee

One peculiar feature of this solution is that the charge parameter $Q$ is related to the integration constants $b$ and $c$ as
\begin{equation}
Q^{2}= 3(1-\gamma) b^2c^2\, .  \label{Qrel}
\end{equation}
This condition is required in \cite{GaoZhang} to generate the analytic solution above. Comparison of (\ref{Qrel}) with (\ref{qrel2}) then also determines the chemical potential as a function of $r_h$ and $b$ as, 
\be
\lab{mu2}
\mu =  \frac{\sqrt{3(1-\g)}}{2} b \, r_h^{3\g-1} (r_h^2 + b^2)^{\half(1-3\g)} \, .
\ee
In passing, we note that the condition (\ref{Qrel}) obscures the physical interpretation in the dual theory. A generic solution to the Einstein-Maxwell-dilaton theory should correspond to a dual field theory that is characterized by three parameters in the grand canonical ensemble: $\Lambda_{QCD}$, $T$ and $\mu$.  If we insist on keeping $\Lambda_{QCD}$---that is related to $b$ as in (\ref{VeV})---and $T$ as the free parameters, then the chemical potential cannot be free. Conversely, we may keep $T$ and $\mu$ free, but then $\Lambda_{QCD}$ will be determined completely. We shall adopt the first option as it is more natural in application to QCD-like theories\footnote{We explain below that existence of a confinement-deconfinement  transition follows from demanding that the high T black-hole and the low T thermal gas  solutions possess the same $\Lambda_{QCD}$.}.  Therefore in this specific model the chemical potential and the mass gap $\Lambda_{QCD}$ will be tied to each other. This unphysical fact is the price one has to pay to work with an analytic solution, which is a crucial technical simplification for the calculations in the next sections.

Finally, we note that the integration constant $c$ can be written in terms of the physical parameters above, using equations (\ref{ceq}), (\ref{qrel2}), (\ref{T2}), (\ref{S2}) and (\ref{mu2}) as 
\be\lab{c2} 
c^2 = 4\pi G T S + Q \mu\, . 
\ee 
This relation clarifies the physical meaning of the integration constant $c$ and it will be useful below when we calculate the energy and the free energy of the solution. 

\subsection{The Thermal Gas solution} 

The TG solution that corresponds to this analytic BH solution is determined by demanding vanishing of the entropy. Noting (\ref{gamma}) and using (\ref{S2}) we learn that TG solution can be obtained from the BH by setting  $r_h=0$.  This is of course expected as the TG solution should follow by sending the horizon to the origin, see e.g. \cite{Longthermo}. Then from (\ref{mu2}) we see that the chemical potential of the TG solution diverges unless $\g\geq 1/3$. This of course does not make sense, thus we further require 
\be\lab{grange1}
 \frac13 \leq \g \leq 1\, .
 \ee
 Then from (\ref{ceq}) we find that the TG solution can be obtained from (\ref{met2}) by setting $c=0$. We present  the thermal gas solution here for completeness, although we will not resort to it in calculations in the next sections: 
  
\begin{equation}\lab{metTG}
d{s}^{2}=-N^{2}(r)\fb^{2}(r)dt^{2}+\frac{r^{2}dr^{2}}{(r^{2}+b^{2})\fb^{2}(r)}%
+(r^{2}+b^{2}){R^{2}(r)}d\Omega _{n-1}^{2}\, ,
\end{equation}
with 
\begin{eqnarray}  \label{grTG}
N^{2}(r) &=&\Gamma ^{-\gamma},  \label{NrTG} \\
\fb^{2}(r) &=&\frac{r^{2}+b^{2}}{l^{2}}\Gamma ^{2\gamma }\, ,
\label{frTG} \\
\Phi(r) &=&\frac{3}{4}\sqrt{\gamma (2-2\gamma)}\log \Gamma\, ,
\label{PhirTG} \\
R^{2}(r) &=&\Gamma ^{\gamma }\, ,  \label{RrTG} \\
\lab{GammaTG}
\Gamma &=&\frac{r^2}{r^{2}+b^{2}}\, .  
\end{eqnarray}
Finally, noting (\ref{Qrel}) that is valid also for the TG solution we learn that {\em the TG solution that corresponds to this BH solution has vanishing charge and vanishing chemical potential:}
\be\lab{vanish} 
Q = \mu = 0, \qquad \textrm{TG}\, ,
\ee
therefore the electromagnetic potential vanishes on the thermal gas solution:
and 
\be\lab{A0TG}
A_t =0,\qquad {\textrm TG} \, .
\ee
This means that the above BH  and the corresponding TG solutions cannot be maintained at the same chemical potential. This is a result of the peculiar condition (\ref{Qrel}). We conclude that one should lok at the {\em canonical ensemble} when studying thermodynamics, rather than the grand canonical ensemble. 


\subsection{The charge and the energy} 

The total charge of the black-hole can be calculated from 
\begin{equation}
Q_{tot}=\frac{1}{4\pi G }\lim_{r \rightarrow \infty }\int d^3x \sqrt{{g_3}} \,N \,Z \,F^{0\nu}n_{\nu}\, ,  \label{Qdef}
\end{equation}
where $N$ is the lapse function in the ADM decomposition, $n_{\mu}$ is the normal vector to the boundary  and $\sqrt{g_3}$ is the volume element of the 3D spatial section. One finds 
\begin{equation}
Q_{tot}=\frac{V_3}{4\pi G}\, Q\, . 
 \lab{Qtot}
 \end{equation}
The gravitational contribution to the mass of the BH solution can be obtained by the Brown-York procedure, that is conveniently reviewed in \cite{Longthermo}. In order to obtain a finite result, it is appropriate to calculate instead the  {\em mass difference}  between the BH and the TG solutions presented above. Employing the expressions presented in \cite{Longthermo} one easily finds the following {\em gravitational contribution} to the mass difference: 
\begin{equation}
\Delta E_G=\frac{3c^2 \,V_3}{16\pi G}\, .  \label{Mass}
\end{equation}
Moreover, using (\ref{c2}) the gravitational mass difference can be expressed as   
\begin{equation}
\Delta E_G=\frac{3\,V_3}{16\pi G} (4\pi G T S + Q \mu)\, .  \label{Mass2}
\end{equation}
There exists also a gauge field contribution to the mass of the black-hole that reads \textbf{add reference}
\be\lab{Es3}
E_{A}^{BH} = \frac{V_3}{4\pi G} \sqrt{g_3} N A_{\n}F^{\m\n} n_{\m} Z(\f)\bigg|^{z=\e}_{z=z_h}  \, .
\ee
On-shell this evaluates to 
\be\lab{Es5}
E_{A}^{BH} = \frac{V_3\, \mu\, Q}{4\pi G} = \mu\, Q_{tot} \, ,
\ee 
Noting the latter contribution is absent in the TG solution, the total mass difference becomes 
\begin{equation}
\Delta E_{tot} =  E_{BH} - E_{TG} =\frac{V_3}{16\pi G} \le(3c^2+ 4\mu\, Q\ri) \, .  \label{Mass3}
\end{equation}

\subsection{Free energy and the Hawking-Page phase transition} 
\lab{Thermodynamics}

Let us now calculate the Gibbs free energy difference between the BH and the TG solutions. One can obtain this from the difference between the on-shell actions employing for example the formula derived in  \cite{Longthermo} and confirms that the result precisely has the form of the Gibbs free energy: 
\begin{equation}
\Delta G_{tot}= \Delta E_G - \mu\, Q_{tot} -T S_{tot} \, .  \label{FE3}
\end{equation}
 Then using (\ref{Mass}) one obtains 
\begin{equation}
\Delta G_{tot}= -\frac{V_3}{16\pi G} c^2  = -\frac14 TS_{tot} - \frac14 \mu\, Q_{tot} \, ,  \label{FE2}
\end{equation}
where we used (\ref{c2}).  

Now we can ask whether there is any phase transition in this system. For this it is useful to express the free energy difference (\ref{FE2})  in terms of $r_h$ and $b$ using  (\ref{S2}), (\ref{T2}), (\ref{mu2}) and (\ref{qrel2}) as 
\begin{equation}
\Delta G_{tot}= -\frac{V_3}{16\pi G} r_h^{6\g-2} (r_h^2+b^2)^{3-3\g}  \, .  \label{FE4}
\end{equation}
This expression can only vanish as $r_h\to 0$ iff $\g\geq \frac13$. Note that this is precisely in the allowed range by (\ref{grange1}). 
Then we can calculate the phase transition temperature $T_c$ by sending $r_h\to 0$ in (\ref{T2}). We find that $T_c$ vanishes for $\g>2/3$ and it diverges for $\g<2/3$. Therefore we only consider the case $\g=2/3$ where $T_c$ turns out to be finite. The value of $T_c$ follows from (\ref{T2}) as 
\be\lab{pt1}
T_{c} = \frac{b}{2\pi} \qquad {\textrm as}\,\,\, r_h\to 0,\qquad {\textrm for}\,\,\,  \g = \frac23\, .
\ee
Recalling that the limit $r_h\to 0$ precisely corresponds to the limit where the mass of the BH vanishes and it becomes the same geometry as the TG background, one is tempted to conclude that this case should correspond to a second order phase transition. However this should be checked by expressing $\Delta G_{tot}$ in terms of $T-T_c$. In order to do this we look at the subleading terms in $T$ and $\Delta G_{tot}$ in $r_h$ as $r_h\to 0$. The result is 
\be\lab{subpt} 
T - T_{c} \approx \frac{3r_h^2}{4\pi b}, \qquad \Delta G_{tot} \approx -\frac{V_3}{16\pi G} b^2 r_h^2\qquad r_h\to 0\, .
\ee  
Therefore we find 
\be\lab{subpt2} 
\Delta G_{tot} \approx -\frac{2 \pi^3 V_3}{ 3G}\, T_{c}^3\, (T-T_{c}), \qquad T\to T_{c}\, ,
\ee  
as $T_c$ is approached from above. This means that this case actually corresponds to a {\em first order phase transition}. 

We note that  this kind of phase transition where the BH horizon marginally traps the singularity at $r_h\to 0$ is of the type considered in \cite{Gursoy-cont-HP}. For further discussion on this we refer to appendix \ref{AppA}.
The Hawking-Page type transition found corresponds to a confinement-deconfinement type transition in the dual field theory\cite{Witten2}.

\section{Anomalous conductivities and fluctuations} 
\lab{main}

Here we calculate the anomalous conductivities for the holographic setup introduced in the previous section.
We follow the methods introduced in  \cite{Landsteiner1} and \cite{HolFlow} where it was applied to the case of ${\cal N}=4$ SYM plasma. 
The conductivities are obtained by finding solutions to the fluctuation equations. 
As an extra check we look at a second derivation, also done in \cite{HolFlow}. 
This derivation uses the fluctuation equations to construct a system of nonlinear differential equations for the Green's functions involved in the conductivities.
We show below that the Green's functions found with the first method above solve these equations. In what follows we only describe the method of calculation and present the final answer, referring all the technical details to Appendices B to F.  

At this point we change notation to make comparison with \cite{Landsteiner1} and \cite{HolFlow} easier. The metric we use is
\begin{equation}
ds^2 = \frac{L^2}{4 u^2 f(u)} du^2 + \frac{\rho_h^2}{L^2 u} \left( - \frac{f(u)}{R(u)^2} dt^2 + R(u)^2 d \vec{x}^2 \right) \, , 
\end{equation}
where $ u = \frac{\rho_h^2}{\rho^2}$ and $\rho^2 = r^2 + b^2$
\begin{equation}\begin{split}
f(u) &= \Gamma^{2\gamma}(u) - \frac{c^2 L^2 u^2}{\rho_h^4} \Gamma^{1-\gamma}\, , \\
R(u)^2 &= \Gamma^\gamma(u) \, , \\
\phi(u) &= \frac{3}{4} \sqrt{\gamma(2 - 2 \gamma)} \log \Gamma(u) \, , \\
\Gamma(u) &= \frac{\rho^2 - b^2}{\rho^2} \, , \\
A_t &= - \frac{\sqrt{3} u b c}{\rho_h^2} \sqrt{1-\gamma} \, , \\
c &= \frac{1}{L} (\rho_h^2 - b^2)^{\frac{1}{2}(3 \gamma - 1)} \rho_h^{\frac{3}{2}(1-\gamma)} \, ,
\end{split}\end{equation}
where we rescaled the gauge field such that the Maxwell term has a factor $\frac{1}{4}$ in front, and otherwise this is just a rewriting of the previous setup, with $u = 0$ corresponding to the boundary and $u =1$ corresponding to the horizon.

A comment is in order here. The calculation we present below relies on the Kubo's linear response theory and therefore the axial currents should have the {\em consistent} form that is related to the {\em covariant} currents by addition of a Chern-Simons current on the boundary, see for example \cite{LandsteinerR}. In the holographic picture this extra piece complicates the calculation and it turns out to use a gauge for the bulk field where $A_t=0$ on the boundary rather than $\mu$ as in the previous section. These two methods were coined the ``formalism A'' and ``formalism B'' in \cite{LandsteinerR}. We shall use formalism B in what follows\footnote{We thank Karl Landsteiner for clarifications on this issue.} in accord with the original holographic calculation of \cite{Landsteiner1}.

The action Eq. \ref{action} discussed in the previous section does not contain the chiral gauge and gravitational anomalies. To introduce them we add the following Chern-Simons type terms\cite{LandsteinerR},

\bea
\lab{CS1} S_{CS} &= \frac{1}{16 \pi G} \int_M d^5 x \sqrt{-g} \frac{\kappa}{3} \epsilon^{MNPQR} A_M F_{NP} F_{QR} \, , \\
\lab{CS2} S_{GCS} &= \frac{1}{16 \pi G} \int_M d^5 x \sqrt{-g} \lambda \epsilon^{MNPQR} A_MR^A{}_{BNP} R^B{}_{AQR}\, , \\
S_{CSK} &=  - \frac{1}{2 \pi G} \int_{\partial M} d^4 x \sqrt{-h} \lambda \epsilon^{MNPQR} n_M A_N K_{PL} D_Q K_R{}^L \, .
\eea
Here $S_{CS}$ is the regular Chern-Simons term and $S_{GCS}$ is the gravitational Chern-Simons term.
The boundary action $S_{CSK}$ needs to be added so that if we do a gauge transformation $A_M \rightarrow A_M + \nabla_M \xi$, the variation of the total action becomes
\begin{equation}
\delta_\xi S = \frac{1}{16 \pi G} \int_{\partial M} \epsilon^{\mu\nu\rho\sigma} \left( \frac{\kappa}{3} \hat{F}_{\mu\nu} \hat{F}_{\rho\sigma} + \lambda \hat{R}^\alpha{}_{\beta\mu\nu} \hat{R}^{\beta}_{\alpha\rho\sigma}  \right) \, .
\end{equation}

To fix $\kappa$ and $\lambda$ we compare with Eq. \ref{axanom1}, where for the single left-handed fermion that we consider, the covariant anomaly has numerical coefficients $a_1 = \frac{1}{96 \pi^2}$ and $a_2 = \frac{1}{768 \pi^2}$, so if we set\footnote{We included a factor of $N_c$ compared to the definitions in \cite{LandsteinerR}  in accord with the generic anomaly equation (\ref{axanom1}).} 
\begin{equation}\label{eq:kl}\begin{split}
\kappa &= - \frac{G N_c}{2 \pi}\, , \\
\lambda &= - \frac{G N_c}{48 \pi} \, ,
\end{split}\end{equation}
we obtain exactly the anomaly equation. Note that we do not at this point include the gluonic contribution to the anomaly, that will be discussed in the next section. 

We also need to add a counter-term action to cancel the divergences at the boundary, (see \cite{Skenderis})
\begin{equation}
S_{ct} = - \frac{1}{8 \pi G} \int d^4 x \sqrt{-h} \left( 3 + \frac{4}{3} \phi^2 \right) \, .
\end{equation}
With the addition of the Chern-Simons terms, the equations of motion become
\begin{equation}\begin{split}\label{eq:eomfull}
0 &= R_{MN} - \frac{1}{2} g_{MN} R + \frac{1}{2} g_{MN} ( \frac{4}{3} (\partial \phi)^2 + V)) + \frac{1}{4} e^{- \frac{4}{3} \alpha \phi} ( \frac{1}{2} g_{MN} F^2 - 2 F_{MP} F_N{}^P ) \\
	&- \frac{4}{3} \partial_M \phi \partial_N \phi - 2 \lambda \epsilon_{L P Q R (M} \nabla_A F^{P L} R^{A}{}_{N)}{}^{QR} \, , \\
0 &= \nabla_N (e^{- \frac{4}{3} \alpha \phi} F^{MN} ) + \epsilon^{MNPQR} ( \kappa F_{N P} F_{Q R} + \lambda R^A{}_{B N P} R^B{}_{A Q R} ) \, , \\
0 &= \frac{8}{3} \frac{1}{\sqrt{-g}} \partial_\mu ( \sqrt{-g} \partial^\mu \phi ) - V(\phi)^\prime + \frac{4\alpha}{12} e^{-\frac{4}{3} \alpha \phi} F^2 \, .
\end{split}\end{equation}
The terms proportional to $\lambda$ or $\kappa$, coming from the new terms in the action, vanish on the background of the previous section. Hence the background introduced in section \ref{Background} still satisfies the equations of motion (\ref{eq:eomfull}).

To compute the Green's functions in the Kubo formulas Eq. (\ref{kubo}) we fluctuate the fields $h_{t \alpha} (r, y)$ and $a_\alpha (r,y)$, where $\alpha$ takes values $x, z$.
To simplify the equations we raise the $\alpha$ index on the metric fluctuations.
Expanding the equations of motion Eq. (\ref{eq:eomfull}) to first order in these perturbations we obtain the fluctuation equations, shown in Appendix \ref{app:flucteq}. 

\subsection{Direct calculation}
\label{directcal}

To calculate the Green's functions directly we use the formalism of \cite{Kaminski}, that we review here.
Second order fluctuations in the action can be written in the form
\begin{equation}
S^{(2)} = \int_M d^4 k \, du \left( \Phi^I_{-k}{}^\prime \tilde{\mathcal{A}}_{IJ} \Phi^J_k{}^\prime + \Phi^I_{-k} \tilde{\mathcal{B}}_{IJ} \Phi^J + \Phi_{-k}^I \tilde{\mathcal{C}}_{IJ} \Phi_k^J\right) \, ,
\end{equation}
where a prime denotes a radial derivative, $\Phi_k$ is a vector of all the (Fourier transformed) fields that we fluctuate and $\tilde{\mathcal{A}}, \tilde{\mathcal{B}}, \tilde{\mathcal{C}}$ are matrices giving the coefficients of the various terms.

Taking this action on-shell, we can write it completely as a boundary term, of the form
\begin{equation}
S^{(2)} = \int_{\partial M} d^4 k_{> 0} \left( \Phi^I_{-k} \mathcal{A}_{IJ} \Phi^J_k{}^\prime + \Phi^I_{-k} \mathcal{B}_{IJ} \Phi^J \right) \, .
\end{equation}
Here we have restricted our integral to momenta  $k_{> 0}$ which are positive in the sense that $\omega > 0$.
$\mathcal{A}$ and $\mathcal{B}$ are then related to $\tilde{\mathcal{A}}$ and $\tilde{\mathcal{B}}$ by  $\mathcal{A} = \frac{1}{2} \tilde{\mathcal{A}}^H$ and $\mathcal{B} = \frac{1}{2} \tilde{\mathcal{B}}^\dagger$, where $\tilde{\mathcal{A}}^H$ denotes the Hermitian part of $\tilde{\mathcal{A}}$.

Now we write $\Phi_k^I(u) = \mathcal{F}^I{}_J(k,u) \phi_k^J$, where $\phi^I$ is the source of $\Phi^I$ and we have a similar relation for $\Phi^I_{-k}$.
We normalize $\mathcal{F}$ so that $\mathcal{F}^I{}_J(k,\Lambda) = \delta^I{}_J$, where $\Lambda$ is the cutoff of our theory. 
This corresponds precisely to requiring our fluctuations to equal the sources at the boundary.
Note that we do not assume the cutoff to be close to the boundary.
Using this definition and normalization of $\mathcal{F}$ we can write the on-shell action as
\begin{equation}
S^{(2)} = \int_{\partial M} d^4 k_{> 0} \phi^I_{-k} 2 \left( \mathcal{A} \mathcal{F}^\prime + \mathcal{B} \right) \phi^J_k |^{\rho_h}_{\Lambda} \, ,
\end{equation}
so that finally,
\begin{equation}\label{greenfct}
G_{IJ}(k) = - 2 \lim_{u\rightarrow \Lambda}  \left( \mathcal{A}(k) (\mathcal{F}(k,r))^\prime + \mathcal{B}(k) \right) \, .
\end{equation}

We find $\mathcal{F}^\prime$ by solving the fluctuation equations.
Apart from the normalization of $\mathcal{F}$ at the boundary we need boundary conditions at the horizon. 
The conditions we need to take are  the $\omega \rightarrow 0$ limit of infalling boundary conditions. This means that the gauge field fluctuations $a_\alpha$ have to be regular at the horizon 
and the metric fluctuations have to vanish (see \cite{Landsteiner2}).

To solve these one has to split the fluctuations into a zeroth order part in momentum and a first order part, where the first order part can again be split up into a term proportional to $\kappa$ and a term proportional to $\lambda$,
\begin{equation}\begin{split}
B_\alpha &= B^{(0)}_\alpha + k \kappa B^{(\kappa)}_\alpha + k \lambda B^{(\lambda)}_\alpha \, , \\
h_t{}^\alpha &= h^{(0)\alpha}_t + k \kappa h_t^{(\kappa)\alpha} + k \lambda h_{ t}^{(\lambda)\alpha} \, .
\end{split}\end{equation}

The zeroth order equation can then be solved by solving the first equation Eq. \ref{eq:fluct1} for $g_{(0) t}^\alpha{}^\prime$ and substituting the result in the second equation Eq. \ref{eq:fluct2}, resulting in an equation of the form
$B^{(0)}_\alpha{}^{\prime\prime\prime} = \chi(u) B^{(0)}_\alpha{}^{\prime\prime}$. 
The solution involves a double integral over the function $\chi(u)$ which cannot be done analytically, but it is divergent at the horizon, so the boundary conditions force it to vanish.
So we obtain an analytic solution for $ B^{(0)}_\alpha$. We plug this back into the first equation Eq. \ref{eq:fluct1} which can then directly be solved for $h^{(0)\alpha}_t$.

The zeroth order solutions then enter in the $\kappa$ and $\lambda$ equations as an inhomogeneous term.
These equations can be solved in the same manner, the only difference being that in this case the boundary conditions do not get rid of the integral.
In this way we obtain an analytic solution involving a double integral, shown in Appendix \ref{app:solutions}.
Note that for the conductivities we only need the asymptotics of the derivatives of the solutions, which we can get from this integral solution fully analytically.

To find the matrices $\mathcal{A}$ and $\mathcal{B}$ we need to look at the second order action.
Our full action can be written as
\begin{equation}
S = S_0 + S_{CS} + S_{GCS} + S_{GH} + S_{CSK} + S_{ct} \, .
\end{equation}
Then the second order actions have the following schematic form (omitting the terms that do not contribute to $\mathcal{A}$ or $\mathcal{B}$),
\begin{equation}\begin{split}
(S_0 + S_{CS} + \partial_u S_{GH} )^{(2)} &= \int_M d^5 x \left( \Phi_{-k}^\prime A_0 \Phi_k^\prime + \Phi_{-k} B_0 \Phi_k^\prime  \right) \, , \\
S^{(2)}_{ct} &= \int_{\partial M} d^4 x \Phi_{-k} B_{ct} \Phi_k   \, , \\
S_{GCS}^{(2)} &= \int_M d^5 x \left( \Phi_{-k}^\prime A_\lambda^1 \Phi_k^\prime + \Phi_{-k} B_\lambda \Phi_k^\prime + \Phi_{-k}^\prime A_\lambda^2 \Phi_k^{\prime\prime}  \right)  \, , \\
S_{CSK}^{(2)} &= \int_{\partial M} d^4 x \left(  \Phi_{-k} A_\lambda^3 \Phi_k^\prime + \Phi_{-k}^\prime A_\lambda^4 \Phi_k^\prime \right)   \, .
\end{split}\end{equation}
All but two of these contributions can be dealt with without problems using the formalism outlined above,
the problematic terms are $A_\lambda^2$ and $A_\lambda^4$.
The reason that these problematic terms are present is that the variational problem is not well defined for the gravitational Chern-Simons term, i.e. if we take the variation of $S_{GCS} + S_{CSK}$ we get
\begin{equation}\label{eq:varprob}
\delta (S_{GCS} + S_{CSK} ) = - \frac{\lambda}{2\pi G} \int_{\partial M} d^4x \sqrt{-h} \epsilon^{mlqr} D_r A_m \delta K_{qv} K_l{}^v \, .
\end{equation}
We deal with these terms by using the fluctuation equations to replace a double derivative on a fluctuation with single derivatives (for the $A_\lambda^4$ term we first have to take the derivative of the entire thing to bring it from the bulk to the boundary).

However this turns out to be insufficient, as in this way the condition
\begin{equation}\label{eq:img}
\frac{d}{du} \left( G - G^\dagger \right) = 0
\end{equation}
is not satisfied. 
We can remedy this as follows, \cite{HolFlow}. 
When we substitute the solutions to the fluctuation equations in Eq. \ref{eq:varprob} it has a term of the form
\begin{equation}
\int_{\partial M} d^4 x \epsilon_{\alpha\beta} \bar{B}_{\alpha -k} \zeta(u,u_c,k) \bar{H}_{\beta k}^\prime \, .
\end{equation}
We take this $\zeta$ as contributing to the $\mathcal{B}$ matrix.
The final matrices are shown in Appendix \ref{app:ab}.

Combining this with the solutions to the fluctuation equations, we find that the condition Eq. \ref{eq:img} is satisfied and we obtain the Green's functions shown in Appendix \ref{app:green}.
Note that at this point we changed our fields (and sources) from $B_\alpha$ and $g_t{}^\alpha$ to $a_\alpha$ and $g^t{}_\alpha$.

Expressed in terms of temperature and chemical potential and using Eq. \ref{eq:kl} we obtain the conductivities in Eq. \ref{eq:cond},
\begin{equation}\label{eq:cond}\begin{split}
\sigma_B &=  \frac{\mu(u_c) }{4 \pi^2} \, , \\
\sigma_V &= - \left( \frac{\mu(u_c)^2}{8\pi^2} +  \frac{T^2}{24} \right) \, , \\
\sigma_B^\epsilon &=  - \left(\frac{\mu(u_c)^2}{8\pi^2} + \frac{T^2}{24} \right) \, , \\
\sigma_V^\epsilon &= \frac{\mu(u_c)^3}{12 \pi^2}+ \frac{1}{12} \mu(u_c) T^2 \, .
\end{split}\end{equation}
There is no explicit $\alpha$-dependence and only the trivial renormalization of $\mu(u_c) = \mu (1- u_c) $.
In particular these results are exactly the same as those in \cite{HolFlow} and agree with \cite{Landsteiner1} and \cite{Landsteiner3}.
Also note the change in sign with respect to (\ref{J2}), this is because here we fluctuate $g^t{}_\alpha$, which corresponds to $T_t{}^\alpha$, while (\ref{J2}) corresponds to $T^{t \alpha}$.

We stress that there is a nontrivial implicit $\alpha$-dependence through the temperature and chemical potential. We refer the reader to Appendix \ref{app:green} where we present  the Green's functions  and their dependence on the parameters of the model $b$, $\rho_h$ and $\alpha$ in detail.

\subsection{Flow equations}
\label{floweqs} 

Following \cite{HolFlow} we now provide another check that the Green's functions found above.
This is only applicable to the $\kappa$ parts however, as it directly uses the stress-energy tensor, and it is not clear how to define this in the presence of the gauge-gravitational anomaly.

The idea is as follows. We have two ways of expressing the current and stress energy tensor. The first is through the Green's functions,
\begin{equation}\begin{split}
\delta J^\alpha_{\text{cons.}} &= G^{xx} \delta^{\alpha\beta} a_\beta + G^{xz} \epsilon^{\alpha\beta} a_\beta + P^{x t} \delta^{\alpha\beta} g^t{}_\beta + P^{z t} \epsilon^{\alpha\beta} g^t{}_\beta \, , \\
\delta T_t{}{}^{\alpha}_{\text{cons.}} &= G^{xx}_\epsilon \delta^{\alpha\beta} a_\beta + G^{xz}_\epsilon \epsilon^{\alpha\beta} a_\beta + P^{x t}_\epsilon \delta^{\alpha\beta} g^t{}_\beta + P^{z t}_\epsilon \epsilon^{\alpha\beta} g^t{}_\beta \, , \\
\end{split}\end{equation}
and the second is as a variation of the action,
\begin{equation}\begin{split}
\delta J^\alpha &= \delta \frac{\delta S}{\delta A_\alpha} = \delta \left( \frac{\sqrt{-g}}{16 \pi G} e^{-\frac{4\alpha}{3} \phi} F^{\alpha r} \right) = \frac{f e^{-\frac{4 \alpha  \phi }{3}} \rho_h^2}{8 \pi  G L^3} \left( a_{\alpha }'+\mu  g^t{}_{\alpha } \right) \, , \\
\delta T_t{}^{\alpha} &= \delta \frac{\delta S}{\delta g^t{}_\alpha} = \delta \left( \frac{\sqrt{-g}}{8 \pi G} \left( K_t{}^\alpha - g_t{}^\alpha K  - \frac{3}{L} h_t{}^\alpha - \frac{L}{2} \hat{G}_t{}^\alpha \right) \right) \\
&= -\frac{f \rho_h^4}{8 \pi  G L^5 R^2 u^2} \left( g^t{}_\alpha \left(u f'+f \left(\frac{2 u R'}{R}-3\right)+3 \sqrt{f}\right)+f u (g^{t }{}_\alpha)^\prime \right)\, . 
\end{split}\end{equation}
Here $K$ is the extrinsic curvature, defined by $K_{\mu\nu} = \frac{1}{2} (\nabla_\mu n_\nu + \nabla_\nu n_\mu)$, where $n_\mu = \sqrt{g_{rr}} dr$ is the radial normal vector, and the stress energy tensor was found in \cite{Kraus}
Note that the $\kappa$ term in the current is absent because we use the covariant current.

Of course these expressions should be the same, and from this we can extract a system of nonlinear differential equations for the correlators.
Equating $(\delta J^\alpha_{\text{cons.}})^\prime = ( \delta J)^\prime$ we get a set of equations involving double and single derivatives of the fields. 
We first get rid of the double derivatives using the fluctuation equations, and then get rid of the single derivatives using directly $\delta J^\alpha_{\text{cons.}} = \delta J^\alpha$.
Then we obtain a system of equations involving only the fields, and the Green's functions and their first derivatives. 
Since the field fluctuations are independent in this context (the two expressions of the current and energy-momentum tensor are equal also off-shell) we find that the coefficients of the fields must vanish individually, 
obtaining a first order coupled system of nonlinear differential equations for the Green's functions, shown in Appendix \ref{app:flow}.

The system is hard to solve directly, but one can easily verify that it is solved by the Green's functions we already found.
More precisely, substituting in in $G^{xx} = P^{xt} = G^{xx}_\epsilon = 0$, as is done in \cite{HolFlow},
the system can be solved (requiring the anomalous conductivities to vanish at the horizon) and we find the same conductivities as with the method above. 
The last Green's function $P^{xt}_\epsilon$ is then solved by 
\begin{equation}
P^{xt}_\epsilon = \frac{f \rho_h^4}{8 \pi  G L^5 R^2 u^2} \left( -u f'-\frac{2 f u R'}{R}+3 f-3 \sqrt{f} \right) \, .
\end{equation}
Hence, the results found above by direct computation agree with the flow equations derived here. This is just an additional check of our results.

\section{Dynamical gauge fields and the Axion}
\lab{gluonsection}

\subsection{Gluonic contribution to the chiral anomaly} 

The conclusion of the last section is that the anomalous conductivities do not get any corrections from interactions in a large class of non-conformal QFTs parametrized by the parameter $\alpha$. More precisely, the form of the conductivities expressed in terms of temperature and the chemical potential is of the same form as the conformal case which corresponds to the choice $\alpha=0$, for which we know that the form of the conductivities is the same as in  free Weyl fermions \cite{Landsteiner1, Landsteiner2, Landsteiner3, Landsteiner4}. 

As discussed in the introduction, this result is expected whenever the strongly interacting QFT can be treated as a hydrodynamical system  where the only hydrodynamical degrees of freedom are the chemical potential $\mu(x)$, the temperature $T(x)$ (or energy density $\epsilon(x)$ and the 4-velocity $u^\mu(x)$ \cite{Yarom}.

 Another piece of information we have is the field theory calculation of \cite{GolkarSon, HouRen}  which shows that the anomalous conductivities can receive quantum corrections through dynamical gluon fields running in the loops. This contribution is obtained in \cite{GolkarSon, HouRen}  by writing the vacuum polarization diagrams $\la A_\mu A_\nu \ra$ in terms of the triangle anomaly diagrams $\la J_\mu G_\nu G_\rho\ra$ where $G_\nu$ are the gluon fields. Therefore we expect such quantum corrections to arise only when the anomaly equation possess these gluonic contribution and read\footnote{We show here the covariant form of the anomalies for the sake of the discussion, as they are more familiar. We also do not include in this discussion the gravitational anomaly for simplicity. We retain it in section B below. } 
 \be\lab{axanom3} 
\6_\mu J^{\m,5} = N_c\,a_1\, \eps^{\m_1\m_2\m_3\m_4}\,\tr\le(F_{\m_1\m_2}F_{\m_3\m_4}\ri)+N_f \, a_3 \, \eps^{\m_1\m_2\m_3\m_4}\,\tr\le(G_{\m_1\m_2}G_{\m_3\m_4}\ri)\, , 
\ee 
instead of just 
 \be\lab{axanom4} 
\6_\mu J^{\m,5} = N_c\,a_1\,  \eps^{\m_1\m_2\m_3\m_4}\,\tr\le(F_{\m_1\m_2}F_{\m_3\m_4}\ri)\, .
\ee 
where $a_1$ and $a_3$ are fixed numerical factors independent of $N_c$ and $N_f$ for the fundamental representation.  In the latter case, when the theory in the strong coupling limit is treated as a hydrodynamical system {\em without the dynamical gluons}, the only hydrodynamical degrees of freedom are  $\mu(x)$, the temperature $T(x)$ and the 4-velocity $u^\mu(x)$, hence one is back to the scenario of \cite{Yarom} and one do not expect quantum corrections. Therefore there is no inconsistency in the results of \cite{GolkarSon, HouRen}  and \cite{Yarom}. The holographic calculation we presented in the previous section also ignores the dynamical gauge field contribution to the anomaly equation hence attaining (\ref{axanom4}) instead of the correct form (\ref{axanom3}). Our calculation fulfills the expectation of \cite{Yarom}, therefore there is neither any inconsistency with the holographic calculation nor with the results of  \cite{GolkarSon, HouRen}. 

However, this logic also suggests a way to obtain such quantum corrections in the holographic description. One simply introduces the necessary bulk degrees of freedom in the GR dual in effect to correct the anomaly equation 
(\ref{axanom4}) and attain (\ref{axanom3}) instead. It is well-known how to obtain such a correction in the holographic picture since the work of Ouyang, Klebanov and Witten \cite{OKW}. It was shown in this paper in the top-down context that such terms arise from the various form fields on the cycles in the internal part of the 10D background. In the particular case of the ${\cal N}=1$ cascading $SU(N+M)\times SU(N)$ gauge theory it arises from the two-form $F_3=dC_2$ on the three cycle in the $T^{1,1}$ geometry. 

The idea was later generalized in \cite{Paredes} in a form suitable for the bottom-up approach we take in this picture.  Instead of reviewing the arguments in \cite{Paredes} let us apply them directly to our case of 5D gravity coupled to the various from fields. Let us first review how the first term in (\ref{axanom3}) arises. In the previous section the anomaly equation arose from the Chern-Simons term   (\ref{CS1})  and (\ref{CS2}). Let us ignore the gravitatonal CS term for the moment, for simplicity. We will put it back in later. In the discussion of \cite{Paredes} the gauge field $A_M$ lives on the flavor 4-branes, and the CS action arises from the Wess-Zumino term of the flavor brane action\footnote{In what follows we denote the holographic coordinate by $u$ in line with the notation of the previous section.}:
\bea
S_{WZ} &=& T_p \int_{\Sigma_{p+1}} C \wedge \text{Str} \exp \left[ i \pi \alpha^\prime \mathcal{F} \right] \nn\\
&=& T_4 \int_M d^5 x  \big\{   i   C_{-1} (u)\wedge \text{Str} \exp [ i 2 \pi \alpha^\prime \mathcal{F} ] |_6 - i \tilde{C}_3 (u) \wedge \text{Str} \exp [ i 2 \pi \alpha^\prime \mathcal{F} ] |_2\nn \\ {}&&-  C_1(u) \wedge \text{Str} \exp [ i 2 \pi \alpha^\prime \mathcal{F} ] |_4 \big\}
\lab{WZ1}
\eea
where, $C$ is the combination of the various form fields $C = \sum_n (-i)^{\frac{p-n+1}{2}}C_n$ and  we specified to flavor 4-branes in the second line. $T_n$ denote the D-brane tensions. The super connection ${\cal F}$ contains the flavor gauge field $A_M$, see \cite{Paredes} for details of the definitions of ${\cal F}$ and the super-trace\footnote{In \cite{Paredes} the super-connection also  includes an open-string tachyon that ${\cal T}$ that we set to zero here. This is expected to be the case in the black-hole backgrounds, see \cite{Jarvinen}. The argument is simple: by symmetry ${\cal T}$ can only be a function of the radial variable $u$. If this function is non-trivial, then one can show by analyzing the equation of motion for ${\cal T}$ that it diverges at the horizon. In \cite{Paredes} it is shown that this gives rise spontaneous breaking of the chiral flavor symmetry, hence a non-trivial value of the chiral condensate. On the other hand we are interested in field theories where the chiral symmetry is restored in the deconfined (hence BH) phase. Then the only sensible solution to the tachyon equation is ${\cal T}=0$ which is indeed always a special solution.} Str.  The super-traces in (\ref{WZ1}) are closed forms that can be written as\footnote{Here we simplify the discussion by first setting the tachyonic contributions to zero, and second by ignoring possible dependence on the dilation in these equations. This is sufficient for the discussion in this section as we only want to discuss modifications due to the Ramond fields in a qualitative manner. The conclusions are unaltered when such contributions are included.}
\bea
\text{Str} \exp [ i 2 \pi \alpha^\prime \mathcal{F} ] |_2 &=&  w_2\, \text{Tr} \left( i F \right) \equiv d \Omega_1 \lab{Str2}\\
\text{Str} \exp [ i 2 \pi \alpha^\prime \mathcal{F} ] |_4 &=& \frac{w_4}{2}\,\text{Tr} \left(     - F \wedge F \right) \equiv d \Omega_3 \lab{Str4}\\
\lab{Str6}
\text{Str} \exp [ i 2 \pi \alpha^\prime \mathcal{F} ] |_6 &=& \frac{w_6}{6} \, \text{Tr}  \big( - i F \wedge F \wedge F  \big)  \equiv d \Omega_5\, ,
\eea
where  $w_2$, $w_6$ and $w_4$ are constants, the first two to be determined by the anomaly equation. 

The first term in (\ref{WZ1}) can therefore be written as $F_0\,\Omega_5$ where $ F_0$ is Poincare dual to a space-filling 5-form $\tilde{F}_5$ which should therefore be constant. This 5-form arises from the gluon $D3$ branes in the decoupling limit, therefore we learn that
$F_0\propto N_c$ for a gauge group $SU(N_c)$. Using the last equation in (\ref{Str6}), choosing the constant $w_6$ appropriately and noting that $\kappa/ G \propto N_c$ (see equation (\ref{eq:kl})) one finds precisely the CS term (\ref{CS1}) with the correct factor of $N_c$ in front. A gauge transformation of $A_M$ then produces a boundary term as the first term in (\ref{axanom3}) hence leads to the holographic representation of the $U(1)$ axial anomaly equation (\ref{axanom4}) \cite{Witten1}. 

The second term in (\ref{WZ1}) involves a three form $\tilde{C}_3$ that has the same degrees of freedom as its Hodge dual $C_0 = \star \tilde{C}_3$. Therefore we can choose to work with the axion\footnote{This is a 5D bulk field, not to be confused with the auxiliary boundary axion in \cite{LandsteinerR}.}  $C_0$. One can easily obtain the action of the axion \cite{Paredes, Iatrakis}. Including the kinetic term for $\tilde{C}_3$ in the bulk action, this dual action takes the form \cite{Iatrakis} 
\be\lab{C0act1} 
S_a = \frac{M_p^3}{2} \int d^5x \sqrt{g} Z_0(\f) \le(dC_0 - N_f \, w_2\, A \ri)^2\, .
\ee 
where $Z_0(\f)$ is some function of the dilation. By symmetry $C_0$ can only be a function of $u$. Here $w_2$ is a constant. For a generic metric of the form 
\be\lab{metgen} 
ds^2 = -g_{tt}(u) dt^2 + g_{uu}(u) du^2 + g_{xx}(u) dx^2_3\, ,  
\ee
and in an arbitrary gauge $A_u$ the solution to (\ref{C0act1}) is 
\be\lab{solC0} 
C_0(u) = \theta + \tilde{\theta} \int^u \sqrt{\frac{g_{uu}}{g_{tt} g_{xx}^3}} Z_0(\f)^{-1} + N_f w_2 \int^u A_u\, ,
\ee  
where $\theta$ and $\tilde{\theta}$ are integration constants. On a BH background then we should require $\tilde{\theta} = 0$ for regularity at the horizon \cite{Longthermo}. On the other hand Maxwell equation for the u-component of the gauge field determines $N_f w_2 A_u = \6_u C_0$. Substituting this in (\ref{solC0}) we then find that the background value of the axion on the BH background is a constant. This constant equals $\theta$ above in the gauge $A_u=0$, however its value changes in  a different gauge. In particular  under $A_u \to A_u + \6_u \lambda$ it transforms as\footnote{$\lambda(u)$ should vanish at the horizon again by regularity.}   
\be\lab{C0tr} \theta \to \theta + N_f w_2 \lambda(0)\, .\ee Now the crucial point is that $C_0$ also couples to the gluon fields that live on the D3 branes \cite{OKW, Paredes}. This coupling can be read off from the Wess-Zumino term on a probe D3 brane as 
\be\lab{WZ2} 
S_{WZ,3} = \frac{T_3}{2} \int d^4x \sqrt{-h} C_0(u) \tr\,  G\wedge G\, = \frac{T_3}{2}\, \theta \int d^4x \sqrt{-h} \tr\,  G\wedge G\, , 
\ee    
where $h$ is the induced metric on the probe brane and this action again follows from a generic form (\ref{WZ1}) where the super-connection ${\cal F}$ is replaced by the gluon field strength $G_{\m\n}$. This is the holographic analog of the theta-term in QCD. Then, as a result of (\ref{C0tr}),  the gauge transformation of $A_u$ in the bulk generates the desired second piece in (\ref{axanom3}) upon appropriate choice of the constant $w_2$. Therefore we learn that {\em in order to include dynamical gauge field contributions to the two-point functions $\la J J \ra$ etc. in (\ref{kubo}) one has to include the axion field $C_0$ in the dual gravitational background.} Now from the axion action (\ref{C0act1}) we see that {\em even though $dC_0=0$ on-shell, this coupling gives rise to a mass term for the bulk gauge field $A$ that is dual to the $U(1)$ axial current.} This is clearly the analog of the Higgsing effect described in \cite{OKW}.   
 
We finally consider the third term\footnote{The role of $C_1$ in the dual field theory is not entirely clear in this setting. It is argued to be dual to the baryon number current in \cite{GK} and indeed $C_1$ may serve as an alternative way to include both the axial and the vector currents in the field theory instead of introducing an additional gauge field on the probe branes as in \cite{Landsteiner3}. Here we will not dwell on the physics of $C_1$  much, but include it in the discussion for completeness. Nevertheless we attain the Wess-Zumino term of the form in (\ref{C1act}) which guarantees absence of any additional anomaly terms on the boundary that would arise form the $C_1$gauge transformations.}  in (\ref{WZ1}), that is of the form $F_2\wedge A\wedge F$. By time translation and rotational  invariance $C_1$ should be of the form $C_1 = C_{1,t}(u) dt + C_{1,u}(u) du$.  Part of the bulk action that contains $C_1$ has the general form 
\be\lab{C1act}  
S_{C1} = \frac{M_p^3}{2} \int d^5x\, \sqrt{g}\, Z_3(\f)\, (dC_1)^2   + T_4 \, N_f\,  w_4\int d^5x\, F_2 \wedge A \wedge F \, ,
\ee 
where $Z_3(\f)$ is a function of the dilation and $w_4$ is a constant. Clearly the last term does not contribute to the equation of motion when $A$ is on-shell. Then one can solve (\ref{C1act}) similarly to (\ref{solC0}) and find
\be\lab{solC1} 
C_1= dt \le(c_1 + c_2 \int^u \sqrt{\frac{g_{uu}g_{tt} }{g_{xx}^3}} Z_3(\f)^{-1}\ri) \, ,
\ee  
where $c_1$ and $c_2$ are integration constants. Note that in contrast to (\ref{solC0}) the constant $c_2$ is not required to vanish by regularity at the horizon. However, one can easily see that  there is no additional term in the anomaly equation (\ref{axanom3}) that would possibly arise from the third term in the Wess-Zumino action (\ref{WZ1}) (that is the second term in (\ref{C1act}) ) because $F_{2}$  only has the $(0u)$ component on-shell. Thus, we managed to satisfy the desired  anomaly  equation (\ref{axanom3}) in the holographic setting. To summarize: {\em the first term in (\ref{axanom3}) arises from the first term in (\ref{WZ1}) and the second term in (\ref{axanom3}) arises from the combination of the second term in (\ref{WZ1}) and (\ref{WZ2}).}  

\subsection{Gluonic contribution to the conductivities} 

Having fixed the Ramond contributions to the bulk action, whose presence is required by the correct anomaly equation (\ref{axanom3}) now we can look at the fluctuations of the various fields to the quadratic order and reconsider the calculation of the two-point function.

We can summarize the findings in the previous subsection by the total action 
\bea
S_{\text{bulk}} &=& M_p^3 N_c^2 \int d^5 x \sqrt{-g} \Bigg( R - \frac{4}{3} (\partial \phi)^2 - \frac{1}{4} Z_1(\f) F^2 - V(\phi) \Bigg) \\
 &&- M_p^3\int d^5 x \sqrt{-g}  \Bigg( \frac{Z_0(\f)}{2} F_1^2 +  \frac{Z_3(\f)}{4} F_2^2\Bigg)\nn \\ 
{}&&+ \int \Bigg( \frac{\tilde{\kappa}}{90}  A\wedge F\wedge F +\frac{\tilde{\lambda}}{30} A \wedge R \wedge R  + \tilde{\xi}\, F_2\wedge A\wedge F  \Bigg) \, ,
\lab{actot}
\eea
where $F_1 = d C_0  - N_f \,w_2 \, A$ and $F_2 = d C_1$ and the constants  $\tilde{\tilde{\kappa}}$, $\tilde{\lambda}$ and $\xi$ are related to $T_4$, $w_4$ and $F_0$ of the previous subsection. We also reinstated the gravitational anomaly term in (\ref{actot}), although this will not play an important role in the discussion below.  For the discussion below it suffices to note that 
\be\lab{scaconsts} 
\tilde{\kappa} \sim \tilde{\lambda} \sim  N_c, \qquad  \tilde{\xi} \sim N_f, \qquad Z_1(\f) \sim \frac{N_f}{N_c},\qquad Z_0(\f)\sim Z_3(\f) \sim 1 \, .
\ee
The background values of the fields are of the form,
\begin{equation}\begin{split}
ds^2 &= - g_{tt}(u) dt^2 + g_{rr}(u) du^2 + g_{xx}(u) d\vec{x}^2 \, , \\
A &= A_t(u) dt \, , \\
C_0 &= \theta \, , \\
C_1 &=  c_1 dt \, ,  
\end{split}\end{equation}

The fluctuation equations are obtained by making the replacement in the action 
\bea
A &\to& A_t(u) dt + a_x(u) e^{-i k_y y} dx +   a_z(u) e^{-i k_y y} dz\nn\\
C_1 &\to& c_1 + \delta C_{1,x}(u) e^{-i k_y y} dx +   \delta C_{1,z}(u) e^{-i k_y y} dz \nn\\
C_0 &\to& \theta + \delta C_0(u)  e^{-i k_y y} \nn\\
g_{MN} &\to& g_{MN} + \delta g_{tx}(u) e^{-i k_y y} + \delta g_{tz}(u) e^{-i k_y y}+\delta g_{yx}(u) e^{-i k_y y} + \delta g_{yz}(u) e^{-i k_y y}\nn
\eea
and expanding it to the quadratic order. Here we first want to discuss three changes in the fluctuation equations 
with respect to the fluctuation equations used in the previous section, that arise from inclusion of the form fields. 
It will be useful define the ratio 
\be\lab{eqx}
x \equiv \frac{N_f}{N_c}\, .
\ee
In the previous section we took $N_f=1$ therefore $x\to 0$ in the 't Hooft limit. Here we would like to discuss the situation in the Veneziano limit 
\be\lab{Veneziano}
N_c\to \infty, \qquad N_f\to \infty, \qquad x \sim 1\, .
\ee
We may still want to consider $x\ll 1$ and expand the solutions in $x$. 

Let us know list the various changes in the fluctuation equations stemming from addition of the form-fields in the game. 
\begin{enumerate}
\item By analyzing scaling of the various terms it is easy to see that the background for $g_{MN}$, $A_M$ and $\f$ is  $\cO(1)$ and for $C_0$ and $C_1$ are $\cO(N_c^{-1})$, therefore one can safely ignore their back reaction on the background without the form fields. 
\item However there is an $\cO(x)$ mass term for the gauge field A that arise from the axion field strength $F_1$. Accordingly the  Maxwell equation for  the {\em background} changes. The maxwell equation now is (ignoring the  
$\cO(N_c^{-1})$ mixing with $C_1$):  
 \be\lab{Maxw}
\6_N\le(\sqrt{-g} Z_1(\f) F^{MN} \ri)=\sqrt{-g} \,x \,w_2 \,Z_0(\f)\, A^M\, .
\ee
Note that the old background with vanishing spatial components $A_i=0$ is still a solution, and we will consider this as a background, however the $A_t$ component necessarily changes because of the $\cO(x)$ mass term, along with the other fields through mixing. We stress that this change is $\cO(x)$: 
\be\lab{change1} 
\{A_t,\f, g_{MN}\} (\textrm{new}) = \{A_t,\f, g_{MN}\} (\textrm{old})  + \cO(x) \, .
\ee 

\item The same mass term affects the fluctuation equations for $a_x$ and $a_z$, the fluctuation equations now become massive. As a result we expect that {\em the anomalous conductivities will get renormalized by order $\cO(x)$.}  This of course has to be confirmed by direct calculation.  
\item There is a mixing between the fluctuations $\delta C_0$ and the longitudinal gauge fluctuations $\delta A_y$ , that we do not consider in this paper. This is of the form $N_f\, k_y\, \delta C_0\, \delta A_y$. Since $C_0 \sim N_c^{-1}$ this mixing affects the fluctuation equation for $\delta A_y$ at order $\cO(x)$. Since this mixing only affects the longitudinal component of the gauge field it should not modify the calculation of the anomalous conductivities.  
\item There exists a mixing between the fluctuation $\delta C_1$ and the gauge fluctuations, that we are interested in, namely $a_x$ and $a_z$. This comes from the $\xi$ term in (\ref{actot}) and is of the form $k_y N_f\, A_t'(u)\, \le(\delta C_{1,x} a_z -\delta C_{1,z} a_y  \ri)$. Again noting that $C_1\sim N_c^{-1}$ this mixing also modifies the gauge fluctuation equations, hence expected to affect the conductivities at order $\cO(x)$.  
\end{enumerate} 
To conclude, we identified {\em three different modifications of the calculation due to presence of the form-fields above, items 2, 3 and 5, which all possibly alter the result at order $\cO(x)$.} This is precisely the same order we expect changes to happen in the dual field theory by addition of the gluonic contribution to the anomaly equation, as comparison of (\ref{axanom3}) with (\ref{axanom4}) shows.


Changes in the fluctuation equations for the metric components is harder to see directly from the action (\ref{actot}). We worked them out explicitly and we present the results below for completeness. Below we present {\em only the new terms that arise from mixing with $\delta C_{1,\mu}$} in the fluctuation equations\footnote{As mentioned above the background itself is affected by $\cO(x)$ on top of the explicit modifications below.}
 for $a_\mu$ and $g_t{}^\mu$  where $\mu = x,z$.  The new terms are shown schematically, i.e. ignore the explicit coefficients, etc. We also show the result only for the $\mu= x$ component, there are similar terms obtained by exchanging $x$ with $z$. We only exhibit dependence on $N_c$ and $x = \frac{N_f}{N_c}$ below, including also the most suppressed contribution in the original equations for comparison, dividing by a factor of $N_c N_f$ everywhere:
\begin{equation}\begin{split}
\text{Maxwell: }& \left( \mathcal{O}(\frac{1}{N_f}) \text{(from $\tilde{\kappa}$, $\tilde{\lambda}$ )}  + \mathcal{O}(1)  \right) + x \left( a_x + A_t g_t{}^x \right) + \frac{1}{N_c} \left( k A_t^\prime \delta C_{1,z} + k A_t^\prime a_z \right) \, , \\
\text{Einstein: }& \left( \mathcal{O}(\frac{1}{N_f}) \text{(from $\tilde{\kappa}$, $\tilde{\lambda}$ )}  + \mathcal{O}(1) \right) + x \left( A_t a_x + A_t^2 g_t{}^x \right) + \frac{1}{N_cN_f} \left( (C_{1,t}^\prime)^2 g_t{}^x + C_{1,t}^\prime \delta C_{1,x}^\prime \right) \, , \\
\text{One-form: }&  A_t^\prime a_z + \frac{1}{N_cN_f} \left( \delta C_{1,x}^{\prime\prime} + \delta C_{1,x}^\prime + C_{1,t}^\prime g_t{}^x \right) \, .
\end{split}\end{equation} 
where prime denotes the derivative with respect to the holographic  coordinate $u$. 
The first terms i the Maxwell and Einstein equations denote the original equations before addition of the p-forms.  The fsecond terms in the Maxwell and Einstein equations are due to the new mass term we described in item 3 above and the first term in the one-form equation is due to the mixing described in item 5. 
We see that the mass term gives corrections of relative order $x$, while the one-form contributions are suppressed with powers of $N_c$.

Also the matrix $\mathcal{B}$ in (\ref{greenfct}) receives an additional contribution of the form
\begin{equation}
\mathcal{B}_{3,1} \propto \frac{ i k \xi}{\sqrt{-g}} C_{1,t} \, , 
\end{equation}
the other matrix entries and the whole $\mathcal{A}$ matrix are unchanged.

\section{Discussion and Outlook} 
\lab{Discussion}

In this paper we extended the holographic calculation of the anomalous conductivities that determine the magnitude of the Chiral Magnetic and the Chiral Vortical effects, to theories that are not conformally invariant. This holographic  calculation for the special case of ${\cal N}=4$ conformal field theory was previously done in \cite{Landsteiner1} and it was found that none of the terms in the anomalous conductivities receive any radiative corrections, hence they strong coupling result fully agrees with the free Weyl fermion limit.  We extend this calculation to non-conformal and confining theories. Our results are expected to be valid in the large $N_c$ and large 't Hooft coupling limit. In particular we considered a bottom-up approach  and focused on a gravitational black-hole background coupled to a non-trivial dilation and gauge field found by Gao and Zhang \cite{GaoZhang}. The background solution is analytic which makes our calculations technically much easier. 

The background has a non-trivial dilation potential that depends on a parameter $\alpha$. In the case $\alpha=0$ it reduces to the cosmological constant term. Therefore these backgrounds can be considered as deformations of ${\cal N}=4$ super Yang Mills theory by a massive operator dual to the dilaton. It turns out this is a deformation by turning on an expectation value for an operator of dimension $\Delta = 2$, hence similar to a mass deformation. Although the scale dimension of the deformation is independent of $\alpha$, the entire background depends on this parameter, and we consider $\alpha$ to be the ``non-conformality'' parameter introducing non-conformality in the system. We showed that for the specific value of $\alpha=2$ the theory admits a confinement-deconfinement type transition that corresponds to a Hawking-Page transition in the gravitational dual. Therefore this is an ideal setting to study the problem of anomalous transport in a strongly coupled but non-conformal theory with a mass gap. 

There is no direct field theory argument that guarantees non-renormalization of the conductivities, and in particular it was shown in \cite{GolkarSon, HouRen} that the coefficient of the $T^2$ term in chiral vortical conductivity does receive radiative corrections. Moreover various lattice studies \cite{lattice} indicate that also the chiral magnetic conductivity can receive corrections. Therefore it is natural to expect radiative corrections especially in theories with an intrinsic mass scale like $\Lambda_{QCD}$ that are in a different universality class than the conformal theories. We computed the chiral magnetic and chiral vortical and the associated heat conductivities in non-conformal theories via the Kubo formulae and yet we find precisely the same form as in the case of ${\cal N} =4$ SYM, indicating that there are no radiative corrections for non-conformal theories either. Our results are presented in equations (\ref{eq:cond}) in section \ref{directcal}. In particular the dependence of the conductivities on the chiral chemical potential and temperature is precisely the same form as in the ${\cal N}= 4$ SYM theory and the free theory limit, showing no explicit dependence on the non-conformality parameter $\alpha$. We emphasize that this result follows in a non-trivial manner: {\em The background, the fluctuation equations, temperature and the chemical potential all exhibit non-trivial and explicit dependence on $\alpha$, yet when the result is expressed in terms of the parameters all the dependence become implicit and the result in terms of physical quantities $\mu$ and $T$ becomes identical to the conformal case.}   We further confirmed these result in an independent manner by working out the ``flow equations'' in section \ref{floweqs} generalizing the results found in \cite{HolFlow}. 

There indeed exists an indirect  argument supporting this non-renormalization property in case the QFT can be described in a hydrodynamic regime and when the contribution of the dynamical gauge fields (gluons) to the anomaly equation is absent \cite{Yarom}. Therefore our results provide both a non-trivial direct check of the intricate arguments presented in \cite{Yarom} and a check of the validity of the holographic calculation in case of non-conformal theories. 

It seems that the resolution of the clash between the direct field theory calculations of  \cite{GolkarSon, HouRen} and \cite{lattice} and the arguments in \cite{Yarom} lies in taking into account the contribution of the gluon fields to the anomaly equation. Motivated by this idea in section \ref{gluonsection} we sought for such corrections in the holographic dual. It is known that such gluonic contributions to the anomaly equation in the holographic picture arise from coupling of the various p-form fields to the flavor and probe branes throughgh the Wess-Zumino terms \cite{OKW, Paredes}. We showed in section \ref{gluonsection} that in the case of the 5D bottom-up approach this role is played by the zero-form $C_0$ (axion) and the one-form $C_1$.  We generalized the fluctuation equations derived in \ref{directcal} by including such form-fields and showed that indeed there exists non-trivial mixing between the gauge and graviton fluctuations with the axion and the one-form fluctuations. It turns out that the calculation is altered in three distinct ways: 1. the background gets corrected through a mass term for the background value of $A_0$, which arises from mixing with the axion, 2. the fluctuation equations for the transverse components $\delta A_x$ and $\delta A_x$ are corrected by the same kind of a mass term, 3. the fluctuation equations further are modified through mixing with the fluctuations of the one-form. All of these corrections turn out to be of order $\cO(N_f/N_c)$ thus only visible in the Veneziano limit $N_c\to \infty$, $N_f\to \infty$ with their ratio fixed. This is also the order that the gluonic contribution to the anomaly equations enter. 

There are various further directions of investigation. First of all the arguments in section \ref{gluonsection} are schematic, and although we present strong arguments for why the conductivities should acquire radiative corrections at order $N_f/N_c$ we have not carried out this calculation in detail to obtain such expected corrections. The reason for this is the technical complications arise in realizing the Veneziano limit holographically. In particular, as mentioned above the background itself receive corrections at this order making the analytic BH solution invalid. We plan to return this calculation in a future work. In particular, it seems most natural to first study the simpler case of ${\cal N}=4$ SYM theory coupled to flavor branes in the Veneziano limit by taking into the back reaction of the D7 or D5 branes on the AdS geometry \cite{Bigazzi, Paredes2, Cotrone}. Without supplying the arguments in section \ref{gluonsection} by detailed calculations we cannot rule out the following possibilities: 1. it may be that the three type of corrections we descrbed in   
section \ref{gluonsection} miraculously cancel each other and yield a result identical to the free limit; 2. It may be that the corrections do change the conductivities but only by correcting the anomaly coefficient which would now include the correction from the gluons.   

Secondly, our calculations are simplified by only considering fermions of one-type of chirality. The true calculation should involve both the left and right gauge fields, although our qualitative results are expected to hold  also after inclusion of the second gauge field \cite{Landsteiner3}. We also plan to consider this issue in our future work. 

A more interesting future study should be to reconsider the arguments in \cite{Yarom} by including the new hydrodynamic degree of freedom that represents the contribution of the gluons to the anomaly of the form $\tr G\wedge G$. This is indeed the operator dual to the axion field considered in section \ref{gluonsection} that seems to  generate the expected modifications. It will be very interesting to study this generalization in the context of the Euclidean partition function considered in \cite{Yarom} in the hydrodynamic limit. Ref. \cite{Kovtun} seems to be a good starting point for such a construction.  

There are various more direct generalizations that involve: the study of the Chiral Separation and the Chiral Vortical Separation effects,  study of the anomalous conductivity two-point functions beyond the IR limit by considering their dependence on the frequency and spatial momentum, studying fluctuations in different channels, e.g. the spin-0 and the spin-2 channels, study of anomalous conductivities also in the low T confining phase, and related to this, searching for discontinuities in the conductivities across the confinement-deconfienment phase transition at $T_c$. 
The Gao-Zhang background we consider in this paper with $\alpha=2$ seems to be a natural starting point for such an investigation.  Finally it is also of interest to carry out such calculations in different holographic backgrounds either in the bottom-up or in the top-down approach.

\section*{Acknowledgments}

We thank  Ioannis Iatrakis, Kristan Jensen, Tigran Kalaydzhyan, Luis Melgar,  Shu Lin, Amos Yarom and especially Karl Landsteiner, Liuba Mazzanti and Francisco Pena-Benitez for interesting discussions.

\appendix

\section{Nature of the phase transition}
\lab{AppA} 

Here we discuss some details of the phase transition discussed in section \ref{Thermodynamics} and compare the situation with the general considerations in \cite{Gursoy-cont-HP} .  It was shown in \cite{Gursoy-cont-HP} that a coordinate independent criterion for the type of ``marginal'' phase transition to happen for the {\em uncharged black-holes} was that the dilaton potential near the transition point behaves as 
\be\lab{mtc}
V(\f) \to e^{\frac43 \f} \le(1 + V_{sub}(\f)\ri),\qquad  \f\to\infty\, ,
\ee
where $V_{sub}$ is a subleading term that determines the order of the transition. One can also formulate this condition in terms of the ``fake super-potential'' as 
\be\lab{mtc1}
W(\f) \to e^{\frac23 \f} \le(1 + W_{sub}(\f)\ri),\qquad  \f\to\infty\, .
\ee
Now one can ask if the special case above with $\gamma=2/3$ satisfies this condition. The dilaton potential in this case reads 
\be\lab{V23}
V(\f) = -4e^{\frac43\f}\le(1 + 2 e^{-2\f}\ri)\, .
\ee
We also note that the ``fake'' super-potential that corresponds to this solution behaves as 
\be\lab{W23}
W(\f) = \pm \frac{3i}{2} e^{\frac23 \f} \le(1 - \frac12 e^{-2 \f} \ri),\qquad  \f\to\infty\, .
\ee
We indeed find that (\ref{V23}) and (\ref{W23}) satisfies (\ref{mtc}) and  (\ref{mtc1}) respectively. The subleading term $\exp(-2\f)$ corresponds to a second order phase transition according to the analysis in \cite{Gursoy-cont-HP}. This seems like a contradiction with the finding above that the transition for $\gamma=2/3$ is first order. The contradiction is resolved however, once one notes that the analysis in  \cite{Gursoy-cont-HP} is only valid for uncharged BHs whereas the solution above is charged.

\section{Fluctuation equations}\label{app:flucteq}
These are the fluctuation equations up to first order in momentum, including both the $\kappa$ and $\lambda$ contributions.
They are written in terms of the parameters $\rho_h$, $v = \frac{b}{\rho_h}$ and $\xi = \frac{\alpha^2-1}{\alpha^2+2}$.
\begin{align}
0 &= 
B_{\alpha }''-\frac{2 u \left(1-v^2\right)^{2 \xi } B_{\alpha }' \left((\xi -1) u v^2+1\right)}{\left(u v^2-1\right) \left(u^2 \left(1-v^2\right)^{2 \xi }-\left(1-u
   v^2\right)^{2 \xi }\right)}+\frac{\left(1-u v^2\right)^{2 \xi } g_t^\alpha{}^\prime}{u^2 \left(1-v^2\right)^{2 \xi }-\left(1-u v^2\right)^{2 \xi }} \nonumber \\
&-\frac{4 i \sqrt{2} k \epsilon _{\alpha \beta } \left(1-v^2\right)^{\xi } \left(1-u v^2\right)^{2 \xi -1}}{\sqrt{1-\xi } v \rho_h \left(u^2 \left(1-v^2\right)^{2 \xi
   }-\left(1-u v^2\right)^{2 \xi }\right)}
   \Big[
   -\kappa  L^2 (\xi -1) v^2 B_{\beta } \left(u v^2-1\right) \label{eq:fluct1} \\
&  + \lambda  u \left(2 \left(\xi ^2-3 \xi +2\right) u^2 v^4+7 (\xi -1) u v^2+3\right) g_t^\beta{}^\prime
   \Big] \, , \nonumber \\
0 &= 
2 (\xi -1) u v^2 \left(1-v^2\right)^{2 \xi } B_{\alpha }' \left(1-u v^2\right)^{-2 \xi -1}+g_t^\alpha{}^{\prime\prime}+\frac{\left(2 \xi  u v^2+1\right)
   g_t^\alpha{}^{\prime}}{u \left(u v^2-1\right)} \nonumber \\
&-\frac{8 i k \lambda  \sqrt{2-2 \xi } u v \epsilon _{\alpha \beta } \left(1-v^2\right)^{\xi }}{\rho_h}
\Big[
u g_t^\beta{}^{\prime\prime}+\frac{\left(2 (\xi +1) u v^2-1\right) g_t^\beta{}^{\prime}}{u v^2-1} \nonumber \\
&-u \left(1-v^2\right)^{2 \xi } B_{\beta }' \left(2 \left(\xi ^2-3 \xi +2\right) u^2 v^4+7 (\xi -1) u v^2+3\right) \left(1-u v^2\right)^{-2 (\xi +1)} \label{eq:fluct2} \\
&+B_{\beta } \left(1-v^2\right)^{2 \xi } \left(1-u v^2\right)^{-2 \xi -3} \big(4 \left(\xi ^2-3 \xi +2\right) u^3 v^6+\left(-6 \xi ^2+25 \xi -19\right) u^2 v^4\\
&-14 (\xi -1) u v^2-3\big) 
\Big] \, \nonumber .
\end{align}

\clearpage
\section{Solutions}\label{app:solutions}
The full solutions to the fluctuation equations can be divided into three terms as follows,
\begin{equation}\begin{split}
B_\alpha &= B_\alpha^{(0)} + i k \epsilon_{\alpha\beta} \left( \kappa B_\beta^{(\kappa)} +  \lambda B_\beta^{(\lambda)} \right) \, , \\
g_t{}^\alpha &= g_t^{(0)}{}^\alpha + i k  \epsilon_{\alpha\beta} \left( \kappa g_t^{(\kappa)}{}^\beta + \lambda g_t^{(\lambda)}{}^\beta \right)  \, .
\end{split}\end{equation}
We will now give all of these terms.

\subsection{Case of $k=0$}
\begin{equation}\begin{split}
B_\alpha^{(0)} &= \bar{B}_{\alpha }+\frac{\left(u-u_c\right) \bar{H}^{\alpha } \left(1-v^2 u_c\right){}^{2 \xi }}{\left(1-v^2 u_c\right){}^{2 \xi }-u_c^2 \left(1-v^2\right)^{2 \xi }} \, , \\
g_t^{(0)}{}^\alpha &= \frac{\left(\left(1-u v^2\right)^{2 \xi }-u^2 \left(1-v^2\right)^{2 \xi }\right) \left(1- u_c v^2\right){}^{2 \xi}}
                              {\left(\left(1-u_c v^2 \right)^{2 \xi }-u_c^2 \left(1-v^2\right)^{2 \xi } \right)  \left(1-u v^2\right)^{2 \xi }} \bar{H}^{\alpha } \, .
\end{split}\end{equation}

\subsection{Finite $\kappa$}
\begin{equation}\begin{split}
B_\beta^{(\kappa)} &=   \int \int \Psi^{(\kappa)}_\beta 
+ \frac{4 i \kappa  k L^2 \sqrt{2-2 \xi } v \epsilon _{\alpha \beta } \left(u_c-1\right) \left(u_c-u\right) \left(1-v^2\right)^{\xi } \bar{B}_{\beta } \left(1-v^2
   u_c\right){}^{2 \xi }}{\rho_h \left(u_c^2 \left(1-v^2\right)^{2 \xi }-\left(1-v^2 u_c\right){}^{2 \xi }\right)}\\
&+\frac{2 i \kappa  k L^2 \sqrt{2-2 \xi } v \epsilon _{\alpha \beta } \left(u_c-1\right){}^2 \left(u_c-u\right) \left(1-v^2\right)^{\xi } \bar{H}^{\beta } \left(1-v^2
   u_c\right){}^{4 \xi }}{\rho_h \left(\left(1-v^2 u_c\right){}^{2 \xi }-u_c^2 \left(1-v^2\right)^{2 \xi }\right){}^2} \, , \\
g_t^{(\kappa)}{}^\beta &= \left(1-u^2 \left(1-v^2\right)^{2 \xi } \left(1-u v^2\right)^{-2 \xi }\right) \int \Psi^{(\kappa)}_\beta \\
&-\frac{4  L^2 \sqrt{2-2 \xi } v  \left(1-v^2\right)^{\xi } \bar{B}_{\beta } \left(1-u v^2\right)^{-2 \xi }}{\rho_h \left(u_c^2
   \left(1-v^2\right)^{2 \xi }-\left(1-v^2 u_c\right){}^{2 \xi }\right)}\times 
\Big(
\left(1-u v^2\right)^{2 \xi } \big((u-1) u_c^2 \left(1-v^2\right)^{2 \xi } \\
&+\left(u_c-u\right) \left(1-v^2 u_c\right){}^{2 \xi }\big) -u^2 \left(u_c-1\right)
   \left(1-v^2\right)^{2 \xi } \left(1-v^2 u_c\right){}^{2 \xi } \Big) \\
&-\frac{2  L^2 \sqrt{2-2 \xi } v  \left(1-v^2\right)^{\xi } \bar{H}^{\beta } \left(1-u v^2\right)^{-2 \xi } \left(1-v^2
   u_c\right){}^{2 \xi }}{\rho_h \left(\left(1-v^2 u_c\right){}^{2 \xi }-u_c^2 \left(1-v^2\right)^{2 \xi }\right){}^2} \times \\
&\Big(
\left(\left(u-u_c\right){}^2 \left(1-u v^2\right)^{2 \xi }-u^2 \left(u_c-1\right){}^2 \left(1-v^2\right)^{2 \xi }\right) \left(1-v^2 u_c\right){}^{2 \xi }\\
&-(u-1) \left(-2 u_c+u+1\right) u_c^2 \left(1-v^2\right)^{2 \xi } \left(1-u v^2\right)^{2 \xi }\Big) \, ,
\end{split}\end{equation}

where integrals are from $u_c$ to $u$ and $\Psi^{(\kappa)}$ is given by
\begin{equation}\begin{split}
\Psi^{(\kappa)}_\beta &= \frac{4 L^2 \sqrt{2-2 \xi } v \left(1-v^2\right)^{\xi } \bar{B}_{\beta } \left(1-u v^2\right)^{2 \xi -1}}{\rho_h \left(\left(1-u v^2\right)^{2 \xi }-u^2
   \left(1-v^2\right)^{2 \xi }\right)^2}
   u \left(1-v^2\right)^{2 \xi } \left(\xi  \left(u^2-1\right) v^2+u+v^2-2\right)\\
   &+\left(1-u v^2\right)^{2 \xi +1} 
   -\frac{4 L^2 \sqrt{2-2 \xi } v \left(1-v^2\right)^{\xi } \bar{H}^{\beta } \left(1-u v^2\right)^{2 \xi -1} \left(1-v^2 u_c\right){}^{2 \xi }}{3 \rho_h \left(\left(1-u
   v^2\right)^{2 \xi }-u^2 \left(1-v^2\right)^{2 \xi }\right)^2 \left(u_c^2 \left(1-v^2\right)^{2 \xi }-\left(1-v^2 u_c\right){}^{2 \xi }\right)} \times \\
&\Big( u \left(1-v^2\right)^{2 \xi } \left(-3 u_c \left(\xi  \left(u^2-1\right) v^2+u+v^2-2\right)+v^2 \left(2 \xi  \left(u^3-1\right)+u^3+2\right)-3\right)\\
&+3 \left(u-u_c\right) \left(1-u v^2\right)^{2 \xi +1}
   \Big) \, .
\end{split}\end{equation}

\subsection{Finite $\lambda$}
\begin{equation}\begin{split}
B^{(\lambda)}_\beta &=  \int \int \Psi^{(\lambda)}_\beta 
-\frac{8 \sqrt{2}}{v \rho_h}
\sqrt{1-\xi } \left(u-u_c\right) \bar{H}^{\beta }
\frac{\left(1-v^2\right)^{\xi -1} \left(1-v^2 u_c\right){}^{2 \xi -1}}{(1-\xi ) \left(\left(1-v^2 u_c\right){}^{2 \xi }-u_c^2 \left(1-v^2\right)^{2 \xi }\right){}^2} \times \\
&\left((\xi -1) v^2+1\right)^2 \left(1-v^2 u_c\right){}^{2 \xi +1}-u_c^3 \left(1-v^2\right)^{2 \xi +1} \left((\xi -1) v^2 u_c+1\right){}^2 \, , \\
g_t^{(\lambda)}{}^\beta &=  \left(1-u^2 \left(1-v^2\right)^{2 \xi } \left(1-u v^2\right)^{-2 \xi }\right) \int \Psi_\beta^{(\lambda)} \\
&+ \frac{8  \sqrt{2-2 \xi }  \left(1-v^2\right)^{3 \xi -1} \bar{H}^{\beta } \left(1-u v^2\right)^{-2 \xi -1} \left(1-v^2
   u_c\right){}^{2 \xi -1}}{(\xi -1) v \rho_h \left(\left(1-v^2 u_c\right){}^{2 \xi }-u_c^2 \left(1-v^2\right)^{2 \xi }\right){}^2} \times \\
   & \Bigl[
   (u-1) u^2  \left(1-v^2 u_c\right){}^{2 \xi }  \big((\xi -1)^2 u (u+1) v^6-(\xi -1) v^4 (\xi +u (\xi +(\xi -1) u-3)-1)\\
   &-2 (\xi -1) (u+1) v^2-1\big)\\
   &+ -(u-1) u^2 v^2 u_c  \left(1-v^2 u_c\right){}^{2 \xi } \big((\xi -1)^2 u (u+1) v^6-(\xi -1) v^4 (\xi +u (\xi +(\xi -1) u-3)-1) \\
   &-2 (\xi -1) (u+1) v^2-1\big)
   + u_c^2 \left(u^3 \left(1-v^2\right)^{2 \xi +1} \left((\xi -1) u v^2+1\right)^2-\left((\xi -1) v^2+1\right)^2 \left(1-u v^2\right)^{2 \xi +1}\right) \\
   &+ u_c^3 \big(\left((\xi -1) v^4 \left((\xi -1) v^2+2\right)+1\right) \left(1-u v^2\right)^{2 \xi +1} \\
   &-u^2 \left(1-v^2\right)^{2 \xi +1} \left((\xi -1) u^2 v^4 \left((\xi -1) u v^2+2\right)+1\right)\big) \\
   &+ 2 (\xi -1) v^2 \left(v^2-1\right) u_c^4 \left(u v^2-1\right) \left(\left(1-u v^2\right)^{2 \xi }-u^2 \left(1-v^2\right)^{2 \xi }\right) \\
   &- (\xi -1)^2 v^4 \left(v^2-1\right) u_c^5 \left(u v^2-1\right) \left(u^2 \left(1-v^2\right)^{2 \xi }-\left(1-u v^2\right)^{2 \xi }\right)
   \Bigr] \, ,
\end{split}\end{equation}
where integrals over $\Psi$ are all from $u_c$ to $u$, and $\Psi^{(\lambda)}$ is given by,

\begin{equation}\begin{split}
\Psi^{(\lambda)}_\beta &= \frac{8 \sqrt{2-2 \xi } u v \left(1-v^2\right)^{3 \xi -1} \bar{B}_{\beta } \left(1-u v^2\right)^{2 (\xi -1)}}{\rho_h \left(\left(1-u v^2\right)^{2 \xi }-u^2   \left(1-v^2\right)^{2 \xi }\right)^2} \times \\
&\Big( 2 (\xi -2) (\xi -1) u^3 \left(v^2-1\right) v^4+7 (\xi -1) u^2 \left(v^2-1\right) v^2+2 (\xi -1) v^2 \left((\xi -1) v^2+2\right)  \\
&-u^3 \left(1-v^2\right)^{2 \xi +1} \left((\xi -1) u v^2 \left(2 u v^2-3\right)-1\right) \left(1-u v^2\right)^{-2 \xi }\\
&+u \left(-2 (\xi -1)^2 v^6-4 (\xi -1) v^4+v^2-3\right)+2 \Big) \\
&+ \frac{8 i \sqrt{2} u \left(1-v^2\right)^{3 \xi -1} \bar{H}^{\beta } \left(1-v^2 u_c\right){}^{2 \xi }}{\sqrt{1-\xi } v \rho_h \left(u v^2-1\right)^2 \left(\left(1-u
   v^2\right)^{2 \xi }-u^2 \left(1-v^2\right)^{2 \xi }\right)^2 \left(u_c^2 \left(1-v^2\right)^{2 \xi }-\left(1-v^2 u_c\right){}^{2 \xi }\right)} \times \\
&\Bigg(
\left(1-u v^2\right)^{2 \xi } \bigg(u \left(v^2 \left(1-2 (\xi -1) v^2 \left((\xi -1) v^2 \left(2 (\xi -1) v^2+5\right)+4\right)\right)-3\right)\\
&+2 \left(2 (\xi -1)  v^2+1\right) \left((\xi -1) v^2+1\right)^2\bigg) \\
&+(\xi -1) u^2 v^2 \left(v^2-1\right) \left(2 u^3 v^2 \left(1-v^2\right)^{2 \xi } \left(-4 \xi +3 (\xi -1) u v^2+6\right)+15 \left(1-u v^2\right)^{2 \xi }\right) \\
&+u^3 \left(v^2-1\right) \left(\left(1-v^2\right)^{2 \xi } \left(-7 (\xi -1) u v^2-1\right)+2 (\xi -1) v^4 \left(1-u v^2\right)^{2 \xi } \left(9 \xi +(\xi -1) (2 \xi -5)
   u v^2-11\right)\right) \\
&\times (\xi -1) v^2 u_c \big(u^3 \left(1-v^2\right)^{2 \xi +1} \left((\xi -1) u v^2 \left(2 u v^2-3\right)-1\right) \\
&-\left(1-u v^2\right)^{2 \xi } \bigg(2 (\xi -2) (\xi -1) u^3 \left(v^2-1\right) v^4+7 (\xi -1) u^2 \left(v^2-1\right) v^2\\
&+u \left(-2 (\xi -1)^2 v^6-4 (\xi -1) v^4+v^2-3\right) +2 \left((\xi -1) v^2+1\right)^2 \bigg)  \Big )\Bigg)
\end{split}\end{equation}

\section{ Matrices $\mathcal{A}$, $\mathcal{B}$}\label{app:ab}
Here we list the matrices $\mathcal{A}$ and $\mathcal{B}$, where everywhere a prefactor of $\frac{\rho_h^4}{16 \pi G L^5}$ is implied but not written.

The matrix $\mathcal{A}$ has as only nonzero components
\begin{equation}\begin{split}
\mathcal{A}_{11} &= \mathcal{A}_{33} = -2 (\xi -1) v^2 \left(1-v^2\right)^{2 \xi } \left(u^2 \left(\frac{v^2-1}{u v^2-1}\right)^{2 \xi }-1\right) \, , \\
\mathcal{A}_{22} &= \mathcal{A}_{44} = \frac{\left(1-u v^2\right)^{2 \xi +1}}{u} \, , \\
\mathcal{A}_{32} &= -\mathcal{A}_{14} = \frac{8 i k \lambda }{\rho_h^2}  \sqrt{2-2 \xi } u^2 v \rho_h \left(1-v^2\right)^{3 \xi } \left((\xi -1) u v^2+1\right) \, , \\
\mathcal{A}_{42} &= - \mathcal{A}_{24} =  \frac{8 i k \lambda }{\rho_h}  \sqrt{2-2 \xi } u v \left(1-v^2\right)^{\xi } \left(1-u v^2\right)^{2 \xi +1} \, . 
\end{split}\end{equation}

The matrix $\mathcal{B}$ can be split up into three parts,  $\mathcal{B}_{CSK}$ coming from $S_{CSK}$, $\mathcal{B}_{CT}$ coming from the counter term and $\mathcal{B}_{AdS+\partial}$ coming from the rest.
$\mathcal{B}_{AdS+\partial}$ is given by 
\begin{equation}
\mathcal{B}_{AdS+\partial} = 
\left(
\begin{array}{cccc}
 0 & -2 v^2 \left(1-v^2\right)^{2 \xi } (\xi -1) & 0 & 0 \\
 0 & -\frac{\left(1-u v^2\right)^{2 \xi } \left(2 u (\xi -1) v^2+3\right)}{u^2} & \mathcal{B}_{AdS+\partial}^{23} & 0 \\
 0 & 0 & 0 & -2 v^2 \left(1-v^2\right)^{2 \xi } (\xi -1) \\
 \mathcal{B}_{AdS+\partial}^{41}  & 0 & 0 & -\frac{\left(1-u v^2\right)^{2 \xi } \left(2 u (\xi -1) v^2+3\right)}{u^2} \\
\end{array}
\right) \, ,
\end{equation}
   
with
\begin{equation}
\mathcal{B}_{AdS+\partial}^{41} = - \mathcal{B}_{AdS+\partial}^{23} = \frac{8 i k \lambda  \sqrt{2-2 \xi } u v \left(1-v^2\right)^{3 \xi } \left(2 \left(\xi ^2-3 \xi +2\right) u^2 v^4+7 (\xi -1) u v^2+3\right)}{\rho_h \left(u v^2-1\right)} \, .
\end{equation}
Note that the component $\mathcal{B}_{AdS+\partial}^{42}$ is also present, but it is proportional to $u - u_c$, so it does not contribute and we do not write it here.

$\mathcal{B}_{CSK}$ has as only nonzero components

\begin{equation}
\mathcal{B}_{CSK}^{41} = - \mathcal{B}_{CSK}^{32} = -\frac{16 i k \lambda  \sqrt{2-2 \xi } u^3 v \left(1-v^2\right)^{5 \xi } \left(1-u v^2\right)^{-2 \xi -1} \left((\xi -1) u v^2+1\right)^2 \left(1-v^2 u_c\right){}^{2
   \xi }}{\rho_h \left(u_c^2 \left(1-v^2\right)^{2 \xi }-\left(1-v^2 u_c\right){}^{2 \xi }\right)} \, .
\end{equation}

Finally $\mathcal{B}_{CT}$ has as only nonzero components
\begin{equation}
\mathcal{B}_{CT}^{22} = \mathcal{B}_{CT}^{44} = \frac{3 \left(1-u v^2\right)^{\frac{1}{3} (7 \xi +2)}}{u^2 \sqrt{\left(1-u v^2\right)^{2 \xi }-u^2 \left(1-v^2\right)^{2 \xi }}} \, .
\end{equation}

Note that there are no $\kappa$ contributions to these matrices. 
Naively there is a contribution to $\mathcal{B}_{31} = - \mathcal{B}_{13}$ but this cancels because we work with the covariant current.

\section{Green's functions}\label{app:green}

Expressed in terms of $\rho_h$, $v = \frac{b}{\rho_h}$ and $\xi= \frac{\alpha ^2-1}{\alpha ^2+2}$, the Green's functions are as follows,
\begin{equation}\begin{split}\label{eq:green}
\langle T_t{}^x T_t{}^x \rangle &= 
-\frac{\rho_h^4 \left(1-v^2 u_c\right){}^{2 \xi } \left(u_c^2 \left(\frac{v^2-1}{v^2 u_c-1}\right){}^{2 \xi }-1\right){}^2}{8 \pi  G L^5 u_c^2} \times \\
&\Big( \frac{2 u_c^2 \left(1-v^2\right)^{2 \xi } \left((\xi -1) v^2 u_c+1\right)}{u_c^2 \left(1-v^2\right)^{2 \xi }-\left(1-v^2 u_c\right){}^{2 \xi }}-2 (\xi -1) v^2
   u_c+\frac{3 \left(1-v^2 u_c\right){}^{\frac{\xi +2}{3}}}{\sqrt{\left(1-v^2 u_c\right){}^{2 \xi }-u_c^2 \left(1-v^2\right)^{2 \xi }}}-3 \Big) \, , \\
\langle J^x J^z \rangle &=  -\frac{i \kappa  k \rho_h}{\sqrt{2} \pi  G L} \left(1-u_c \right) \sqrt{1-\xi } v \left(1-v^2\right)^{\xi } \, , \\
\langle J^x T_t{}^z \rangle  &=
\frac{i \kappa  k v^2 \rho_h^2}{2 \pi  G L^2} 
(1 - \xi) \left(1-u_c\right){}^2 \left(1-v^2\right)^{2 \xi } +
\frac{2 i k \lambda  \rho_h^2}{\pi  G L^4}  
\left(1-v^2\right)^{2 \xi -1} \left((\xi -1) v^2+1\right)^2    \, , \\
\langle T_t{}^x J^z \rangle  &= 
\frac{i \kappa  k v^2 \rho_h^2}{2 \pi  G L^2} 
(1 - \xi) \left(1-u_c\right){}^2 \left(1-v^2\right)^{2 \xi }  +
\frac{2 i k \lambda  \rho_h^2}{\pi  G L^4}  
\left(1-v^2\right)^{2 \xi -1} \left((\xi -1) v^2+1\right)^2     \, , \\
\langle T_t{}^x T_t{}^z \rangle  &= 
-\frac{i \sqrt{2} \kappa  k v^3 \rho_h^3}{3 \pi  G L^3}
\left(1-u_c\right){}^3
(1-\xi )^{3/2} \left(1-v^2\right)^{3 \xi } \\
 &- \frac{4 i k \lambda  \rho_h^3}{\pi  G L^5}  \sqrt{2-2 \xi } (1-u_c) v \left(1-v^2\right)^{3 \xi -1} \left((\xi -1) v^2+1\right)^2  \, ,
\end{split}\end{equation}
with other components vanishing.

\section{Flow equations}\label{app:flow}

\begin{equation}\begin{split}
G_{\text{xx}}' &= -\frac{8 \pi  G L^5 R^2 u \left(P_{\text{zt}} G_{\text{xz}}^{\epsilon }-P_{\text{xt}} G_{\text{xx}}^{\epsilon }\right)}{f^2 \rho_h^4}-\frac{8 \pi  G L^3 e^{\frac{4 \alpha \phi }{3}} \left(G_{\text{xx}}^2-G_{\text{xz}}^2\right)}{f \rho_h^2} \, , \\
G_{\text{xz}}' &= \frac{8 \pi  G L^5 R^2 u \left(P_{\text{xt}} G_{\text{xz}}^{\epsilon }+P_{\text{zt}} G_{\text{xx}}^{\epsilon }\right)}{f^2 \rho_h^4}-\frac{16 \pi  G L^3 e^{\frac{4 \alpha \phi }{3}} G_{\text{xx}} G_{\text{xz}}}{f \rho_h^2}+\frac{i \kappa  k \mu }{2 \pi  G} \, , \\
P_{\text{xt}}' &=  -\frac{8 \pi  G L^5 R^2 u P_{\text{zt}}^{\epsilon}P_{\text{zt}}}{f^2 \rho_h^4}-G_{\text{xx}} \left(\frac{8 \pi  G L^3 e^{\frac{4 \alpha  \phi }{3}} P_{\text{xt}}}{f \rho_h^2}-\mu \right)+\frac{8 \pi  G L^3 e^{\frac{4 \alpha  \phi }{3}} G_{\text{xz}} P_{\text{zt}}}{f \rho_h^2}-							P_{\text{xt}} \Pi _{\text{xt}}^{\epsilon }\, , \\
P_{\text{zt}}' &= \frac{8 \pi  G L^5 R^2 u P_{\text{xt}} P_{\text{zt}}^{\epsilon }}{f^2 \rho_h^4}-G_{\text{xz}} \left(\frac{8 \pi  G L^3 e^{\frac{4 \alpha  \phi }{3}} P_{\text{xt}}}{f \rho_h^2}-\mu \right)-\frac{8 \pi  G L^3 e^{\frac{4 \alpha  \phi }{3}} G_{\text{xx}} P_{\text{zt}}}{f \rho_h^2}-						P_{\text{zt}} \Pi _{\text{xt}}^{\epsilon } \, , \\
\left(G_{\text{xx}}^{\epsilon }\right)' &= -\frac{8 \pi  G L^5 R^2 u G_{\text{xz}}^{\epsilon } P_{\text{zt}}^{\epsilon }}{f^2 \rho_h^4}-G_{\text{xx}} \left(\frac{8 \pi  G L^3 e^{\frac{4 \alpha  \phi }{3}}
   G_{\text{xx}}^{\epsilon }}{f \rho_h^2}-\mu \right)+\frac{8 \pi  G L^3 e^{\frac{4 \alpha  \phi }{3}} G_{\text{xz}}^{\epsilon}G_{\text{xz}}}{f \rho_h^2}-G_{\text{xx}}^{\epsilon } \Pi
   _{\text{xt}}^{\epsilon }\, , \\
\left(G_{\text{xz}}^{\epsilon }\right)' &= \frac{8 \pi  G L^5 R^2 u G_{\text{xx}}^{\epsilon } P_{\text{zt}}^{\epsilon }}{f^2 \rho_h^4}-G_{\text{xz}} \left(\frac{8 \pi  G L^3 e^{\frac{4 \alpha  \phi }{3}}
   G_{\text{xx}}^{\epsilon }}{f \rho_h^2}-\mu \right)-\frac{8 \pi  G L^3 e^{\frac{4 \alpha  \phi }{3}} G_{\text{xx}} G_{\text{xz}}^{\epsilon }}{f
   \rho_h^2}-G_{\text{xz}}^{\epsilon } \Pi _{\text{xt}}^{\epsilon } , \\
\left(P_{\text{xt}}^{\epsilon }\right)' &=  -\frac{8 \pi  G L^5 R^2 u \left(P_{\text{zt}}^{2 \epsilon }-P_{\text{xt}}^{2 \epsilon }\right)}{f^2 \rho_h^4}-P_{\text{xt}}^{\epsilon } \left(-\frac{2 f'}{f}-\frac{4 R'}{R}+\frac{6}{u}\right)-P_{\text{xt}} \left(\frac{8 \pi  G L^3 e^{\frac{4 \alpha  \phi }{3}} G_{\text{xx}}^{\epsilon }}{f \rho_h^2}-\mu   \right)\\
&+\frac{8 \pi  G L^3 e^{\frac{4 \alpha  \phi }{3}} P_{\text{zt}} G_{\text{xz}}^{\epsilon }}{f \rho_h^2} +\mu  G_{\text{xx}}^{\epsilon} \\
						     &   -\frac{f \rho_h^4 \left(R u \left(R u f''-f' \left(2 u R'+R\right)\right)+f \left(-3 R^2-10 u^2 \left(R'\right)^2+2 R u \left(u R''+8 R'\right)\right)\right)}{8 \pi  G L^5 R^4 u^3}  \, , \\
\left(P_{\text{zt}}^{\epsilon }\right)' &= \frac{16 \pi  G L^5 R^2 u P_{\text{xt}}^{\epsilon } P_{\text{zt}}^{\epsilon }}{f^2 \rho_h^4}-P_{\text{zt}}^{\epsilon } \left(-\frac{2 f'}{f}-\frac{4
   R'}{R}+\frac{6}{u}\right)-\frac{8 \pi  G L^3 e^{\frac{4 \alpha  \phi }{3}} P_{\text{xt}} G_{\text{xz}}^{\epsilon }}{f \rho_h^2}\\
   &-P_{\text{zt}} \left(\frac{8 \pi  G L^3 e^{\frac{4 \alpha  \phi }{3}} G_{\text{xx}}^{\epsilon }}{f \rho_h^2}-\mu \right)
   						 +\mu  G_{\text{xz}}^{\epsilon }  \, , \\
\end{split}\end{equation}
with
\begin{equation}
\Pi_{xt}^\epsilon = \left(\frac{2 f^2 u \rho_h^4 R'-3 f^2 R \rho_h^4+f R u f' \rho_h^4}{f^2 R u \rho_h^4}+\frac{8 \pi  G L^5 R^2 u P_{\text{xt}}^{\epsilon }}{f^2 \rho_h^4}\right) \, .
\end{equation}

\end{document}